\documentclass[10pt, twoside, journal, compsoc]{IEEEtran}

\ifCLASSINFOpdf
  \usepackage[pdftex]{graphicx}
  \graphicspath{{../pdf/}{../jpeg/}}
  \DeclareGraphicsExtensions{.pdf,.jpeg,.png}
\else
  \usepackage[dvips]{graphicx}
  \graphicspath{{../eps/}}
  \DeclareGraphicsExtensions{.eps}
\fi
\usepackage{algorithm}
\usepackage{algpseudocode}
\ifCLASSOPTIONcompsoc
  \usepackage[caption=false,font=footnotesize,labelfont=sf,textfont=sf]{subfig}
\else
  \usepackage[caption=false,font=footnotesize]{subfig}
\fi
\usepackage{listings}
\usepackage[dvipsnames]{xcolor}
\usepackage{float}
\usepackage{pifont}
\ifCLASSOPTIONcompsoc
  \usepackage[nocompress]{cite}
\else
  \usepackage{cite}
\fi
\usepackage{url}
\usepackage{color, colortbl}
\usepackage{xspace}
\usepackage{booktabs}
\usepackage[american]{babel}
\usepackage{listings}
\usepackage{flushend}
\usepackage{multirow}
\usepackage{amsmath}
\usepackage{setspace}
\usepackage{soul}

\makeatletter
\def\therule{\makebox[\algorithmicindent][l]{\hspace*{.5em}\vrule height .75\baselineskip depth .25\baselineskip}}%

\newtoks\therules
\therules={}
\def\appendto#1#2{\expandafter#1\expandafter{\the#1#2}}
\def\gobblefirst#1{
  #1\expandafter\expandafter\expandafter{\expandafter\@gobble\the#1}}%
\def\LState{\State\unskip\the\therules}
\def\pushindent{\appendto\therules\therule}%
\def\popindent{\gobblefirst\therules}%
\def\printindent{\unskip\the\therules}%
\def\printandpush{\printindent\pushindent}%
\def\popandprint{\popindent\printindent}%

\algdef{SE}[WHILE]{While}{EndWhile}[1]
  {\printandpush\algorithmicwhile\ #1\ \algorithmicdo}
  {\popandprint\algorithmicend\ \algorithmicwhile}%
\algdef{SE}[FOR]{For}{EndFor}[1]
  {\printandpush\algorithmicfor\ #1\ \algorithmicdo}
  {\popandprint\algorithmicend\ \algorithmicfor}%
\algdef{S}[FOR]{ForAll}[1]
  {\printindent\algorithmicforall\ #1\ \algorithmicdo}%
\algdef{SE}[LOOP]{Loop}{EndLoop}
  {\printandpush\algorithmicloop}
  {\popandprint\algorithmicend\ \algorithmicloop}%
\algdef{SE}[REPEAT]{Repeat}{Until}
  {\printandpush\algorithmicrepeat}[1]
  {\popandprint\algorithmicuntil\ #1}%
\algdef{SE}[IF]{If}{EndIf}[1]
  {\printandpush\algorithmicif\ #1\ \algorithmicthen}
  {\popandprint\algorithmicend\ \algorithmicif}%
\algdef{C}[IF]{IF}{ElsIf}[1]
  {\popandprint\pushindent\algorithmicelse\ \algorithmicif\ #1\ \algorithmicthen}%
\algdef{Ce}[ELSE]{IF}{Else}{EndIf}
  {\popandprint\pushindent\algorithmicelse}%
\algdef{SE}[PROCEDURE]{Procedure}{EndProcedure}[2]
   {\printandpush\algorithmicprocedure\ \textproc{#1}\ifthenelse{\equal{#2}{}}{}{(#2)}}%
   {\popandprint\algorithmicend\ \algorithmicprocedure}%
\algdef{SE}[FUNCTION]{Function}{EndFunction}[2]
   {\printandpush\algorithmicfunction\ \textproc{#1}\ifthenelse{\equal{#2}{}}{}{(#2)}}%
   {\popandprint\algorithmicend\ \algorithmicfunction}%

\newcommand\etal{\emph{et al.}\xspace}
\newcommand\Modela{\emph{Liu et al.}\xspace}
\newcommand\Modelb{\emph{Werkhoven et al.}\xspace}

\newcommand\cparagraph[1]{\vspace{1.5mm}\noindent \textbf{#1}}

\definecolor{mygreen}{rgb}{0,0.6,0}
\newcommand{\SVMs} {\texttt{SVMs}\xspace}
\newcommand{\SVM} {\texttt{SVM}\xspace}
\newcommand{\DCT} {\texttt{DCT}\xspace}
\newcommand{\MLP} {\texttt{MLP}\xspace}
\newcommand{\PCA} {\texttt{PCA}\xspace}

\newcommand{\ANN} {\texttt{ANN}\xspace}
\newcommand{\FA} {\texttt{FA}\xspace}
\newcommand{\SVMC} {\texttt{SVM-classifier}\xspace}

\newcommand{\hStreams} {\textsc{hStreams}\xspace}

\definecolor{Gray}{gray}{0.9}

\makeatother

\begin{document}
\title{Optimizing Streaming Parallelism on Heterogeneous Many-Core Architectures: A Machine Learning Based Approach}

\author{Peng~Zhang,
        Jianbin~Fang,
        Canqun~Yang,
        Chun~Huang,
        Tao~Tang,
        Zheng~Wang%

\thanks{
}

    \IEEEcompsocitemizethanks{
        \IEEEcompsocthanksitem Peng Zhang, Jianbin Fang, Canqun Yang, Chun Huang, and Tao Tang are with
         National University of Defense Technology, China.\protect\\
        E-mail: \{zhangpeng13a, j.fang, canqun, chunhuang\}@nudt.edu.cn
        \IEEEcompsocthanksitem Zheng Wang is with University of Leeds, United Kingdom.\protect\\
        E-mail: z.wang5@leeds.ac.uk}%
}

\IEEEtitleabstractindextext{
	\begin{abstract}


    As many-core accelerators keep integrating more processing units, it becomes increasingly more difficult for a
    parallel application to make effective use of all available resources. An effective way for improving hardware
    utilization is to exploit spatial and temporal sharing of the heterogeneous processing units by multiplexing
    computation and communication tasks -- a strategy known as heterogeneous streaming. Achieving effective
    heterogeneous streaming requires carefully partitioning hardware among tasks, and matching the granularity of
    task parallelism to the resource partition. However, finding the right resource partitioning and task granularity
    is extremely challenging, because there is a large number of possible solutions and the optimal solution varies
    across programs and datasets. This article presents an automatic approach to quickly derive a good solution for
    hardware resource partition and task granularity for task-based parallel applications on heterogeneous many-core
    architectures.  Our approach employs a performance model to estimate the resulting performance of the target
    application under a given resource partition and task granularity configuration. The model is used as a utility
    to quickly search for a good configuration at runtime. Instead of hand-crafting an analytical model that requires
    expert insights into low-level hardware details, we employ machine learning techniques to automatically learn it.
    We achieve this by first learning a predictive model offline using training programs. The learnt model can then
    be used to predict the performance of any unseen program at runtime. We apply our approach to 39 representative
    parallel applications and evaluate it on two representative heterogeneous many-core platforms:  a CPU-XeonPhi
    platform and a CPU-GPU platform. Compared to the single-stream version, our approach achieves, on average, a 1.6x and 1.1x speedup on the XeonPhi and
    the GPU platform, respectively. These results translate to over 93\% of the performance delivered by a
    theoretically perfect predictor.

\end{abstract}

\begin{IEEEkeywords}
Heterogeneous computing; Parallelism; Performance Tuning; Machine learning
\end{IEEEkeywords}}


\maketitle

\IEEEdisplaynontitleabstractindextext

\section{Introduction}\label{sec:introduction}
Heterogeneous many-cores, as representative by GPGPUs and Intel's XeonPhi, are widely used for accelerating parallel
applications~\cite{citeulike:2767438, DBLP:conf/sc/LiLKVWHMS17, DBLP:journals/computing/ChenFTY17}. As users demand higher performance,
many-core accelerators have become more powerful by providing more and more processing units. While the abundant computing resources offer
the potential for higher performance,  it becomes harder for a parallel application to utilize all the available computing
resources~\cite{citeulike:14070672, DBLP:conf/icpp/FangVS11}. As a result, many parallel applications fail to fully unlock the performance
potential of a many-core accelerator.

One way for improving heterogeneous many-core utilization is to exploit spatial and temporal sharing of processing resources. This
strategy is also known as \textit{heterogeneous streaming}~\cite{DBLP:conf/ipps/NewburnBWCPDSBL16}. The idea
is to exploit the computation and communication independency of task parallelism to improve hardware utilization. It works by partitioning
the processor cores to allow independent communication and computation tasks (i.e. streams) to run concurrently on different hardware
resources, which effectively overlaps the concurrent kernel execution with data movements. Representative heterogeneous streaming
implementations include CUDA Streams~\cite{tr:cuda:best}, OpenCL Command Queues~\cite{website:opencl_ref}, and Intel heterogeneous
streams library (\hStreams)~\cite{tr:hstreams:arch, DBLP:conf/ipps/NewburnBWCPDSBL16}. These implementations allow a parallel program to spawn more than one
stream (or pipeline) so that the data movement stage of one pipeline overlaps the kernel execution stage of another.

Prior work on heterogeneous streaming mainly targets GPUs~\cite{citeulike:13920334, citeulike:9715521, citeulike:13920353}. Compared to GPU
implementations, OS-enabled coprocessors, like the Intel XeonPhi, provides some unique features that are currently unavailable on the GPU.
For example, besides specifying the number of streams, developers can explicitly map streams to different groups of cores on XeonPhi to
control the number of cores of each hardware partition. This parameter is not exposed to programmers on GPUs, making previous work on
GPU-based parallel streaming optimizations infeasible to fully exploit Xeon-Phi-like many-core accelerators (see also
Section~\ref{sec:alt}). On the other hand, ample evidence is showing that choosing the right stream configuration, i.e., the number
of processor core partitions and the number of concurrent tasks of a multi-stream application, values, has a significant impact the
application's performance on many-core architectures~\cite{DBLP:conf/npc/LiFTCY16, DBLP:journals/ppl/FangZLTCCY16,
DBLP:conf/ipps/LiFTCCY16}. However, attempting to find the optimal values through exhaustive profiling would be ineffective, because the
range of the possible values for the two parameters is huge. What we need is a technique that automatically determines the optimal stream
configuration for any streamed application in a fast manner.

This article presents a novel approach to determine the right number of processor core partitions and tasks for heterogeneous streams,
targeting heterogeneous many-core architectures. Our key insight is to use a performance model to quickly search for the optimal stream
configuration. The performance model estimates the resulting performance of the target streamed application when it runs under a given
stream configuration. If the prediction can be performed quickly with low overhead, we can then quickly explore a large configuration
space. Instead of hand-crafting the performance model that requires human modification whenever the architecture evolves (i.e., when
the number and types of cores change), we employ machine learning techniques to automatically construct a predictive model. Our predictor
is first trained \emph{off-line}. Then, using code and dynamic runtime features of the program, the model predicts performance for a
\emph{new}, \emph{unseen} program under a given stream configuration.

Our prior  work~\cite{ipdps18} develops a machine learning based classifier to predict the optimal stream configuration. However, this
approach can only choose from a limited set of configurations seen during the training phase. Unlike a classification-based approach, the
approach presented in the article allows us to explore a larger number of stream configurations (including those that are not seen during
the training phase) with negligible runtime
overhead. This advantage significantly improves the generalization ability of the proposed approach (Section~\ref{sec_mlstream_modeling}).

Due to the newness of heterogeneous streaming execution model, there are very few multi-stream benchmarks available. To evaluate our
approach on a wide range of applications, we have developed a compiler-based tool to automatically translate standard OpenMP benchmarks
into their streamed variants
for the backends of XeonPhi and GPU architectures (Section~\ref{sec_omp2hstream}).
With the help of this code generator, we can apply our approach to 39 parallel benchmarks.
We argue that this tool can help generate more streamed code and thus is an added value to the community.

We evaluate our approach on two representative heterogeneous many-core platforms: a 57-core Intel XeonPhi
and an NVIDIA 1080Ti GPU platforms. We achieve, on average, a 1.6x and 1.1x speedup over the single-stream execution on the XeonPhi and
the GPU platforms, respectively. This translates to over 93\% of the best available performance.

%
%

The core contribution of this paper is a novel machine-learning-guided approach for automatically determining the optimal stream
configuration on heterogeneous many-cores. We show that our approach delivers good performance across benchmarks and heterogeneous
many-core platforms. While we do not seek to advance the machine learning algorithm itself, our work shows how machine learning can be used
to address the challenging problem of tuning fine-grained streaming parallelism  on heterogeneous many-core architectures. In this work, we
demonstrate the usefulness of our approach on XeonPhi and an NVIDIA GPU, but our approach is equally
applicable on other heterogeneous platforms like AMD GPUs.

\section{Background and Overview} \label{sec_mlstream_motivation}
In this section, we first give a brief introduction of heterogeneous streaming; we then define the scope of this work, before motivating
the need of our scheme and providing an overview of our approach.

\subsection{Heterogeneous Streaming} \label{subsec:multiple:streams}
The idea of heterogeneous streaming is to exploit spatial and temporal sharing of computing resources to utilize the hardware resources to
improve application performance.


\cparagraph{Spatial Sharing.} Modern many-core accelerators offer a large number of processing units. Since many applications cannot fully
utilize all the cores at a time, we can partition the computing units into multiple groups to concurrently execute multiple tasks. In this
way, the computing resource is spatially shared across concurrently-running application tasks. The key to spatial sharing is to determine
the right number of partitions, because over-provisioning of processing units would waste computing resources but under-provisioning would
lead to slowed down performance.

\cparagraph{Temporal Sharing.} Code written for heterogeneous computing devices typically consists of several stages, such as host device
communication and computation. Using temporal sharing, one can overlap some of these stages to exploit pipeline parallelism to improve
performance by overlapping the host-device communication and kernel execution.


\lstset{}
\begin{figure}
	\centering %
		\noindent\mbox{\parbox{\columnwidth}{%
\begin{lstlisting}[label=subfig:source_in]
//setting the partition-size and task granularity
hStreams_app_init(partition_size,streams_p_part);

//stream queue id
stream_id = 0;
for(...){
  //enquque host-device transfer to current stream
  hStreams_app_xfer_memory(,,, stream_id, HSTR_SRC_TO_SINK,...);
  ...
  //enqueue computation to the current stream
  hStreams_EnqueueCompute(stream_id, "kernel1", ...);
  ...
  //move to the next stream
  stream_id = (stream_id++) % MAX_STR;
}
//transfer data back to host
hStreams_app_xfer_memory(,,, HSTR_SINK_TO_SRC,...);
\end{lstlisting}%
		}}
    \vspace{-3mm}
	\caption{Heterogeneous streaming using \hStreams as an example.}%
	\label{fig:example_code}%
\end{figure}

\begin{figure*}[!t]
\centering
\subfloat[\texttt{binomial}]{\label{fig_motivation_partition_cf}\includegraphics[width=0.48\textwidth]{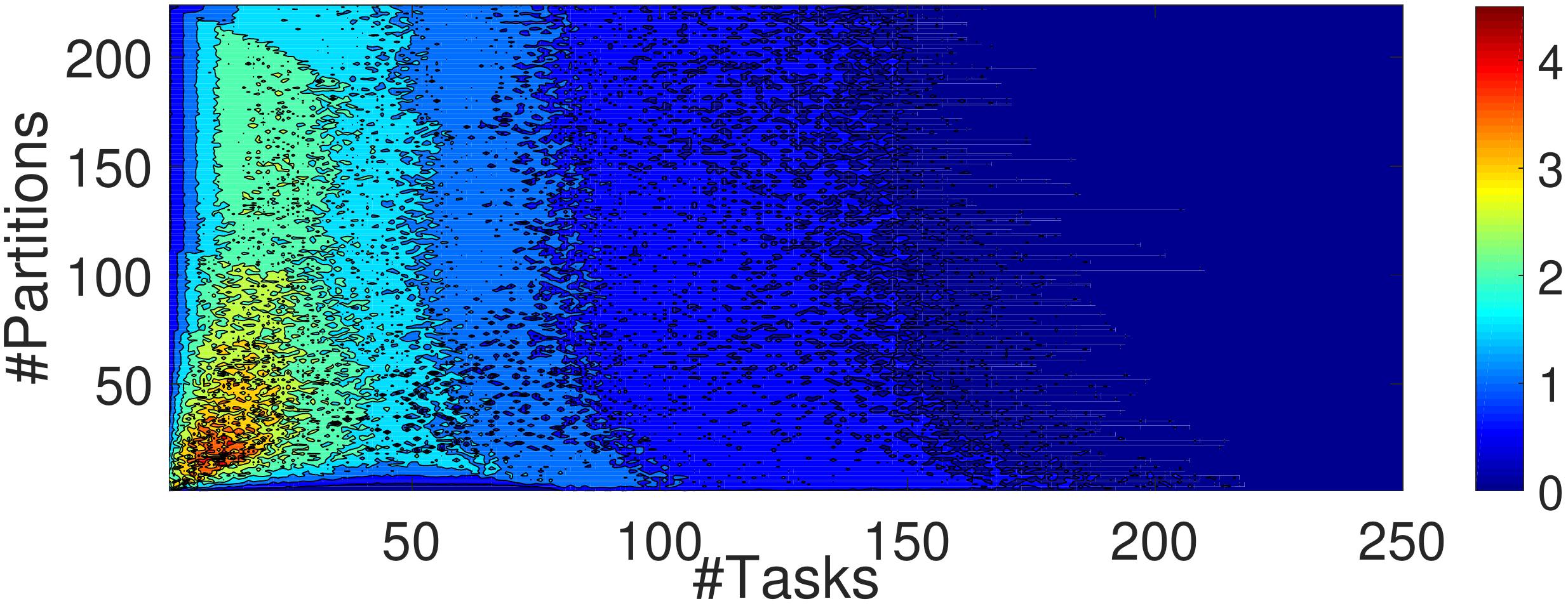}}
\hfil
\subfloat[\texttt{prefixsum}]{\label{fig_motivation_partition_mm}\includegraphics[width=0.48\textwidth]{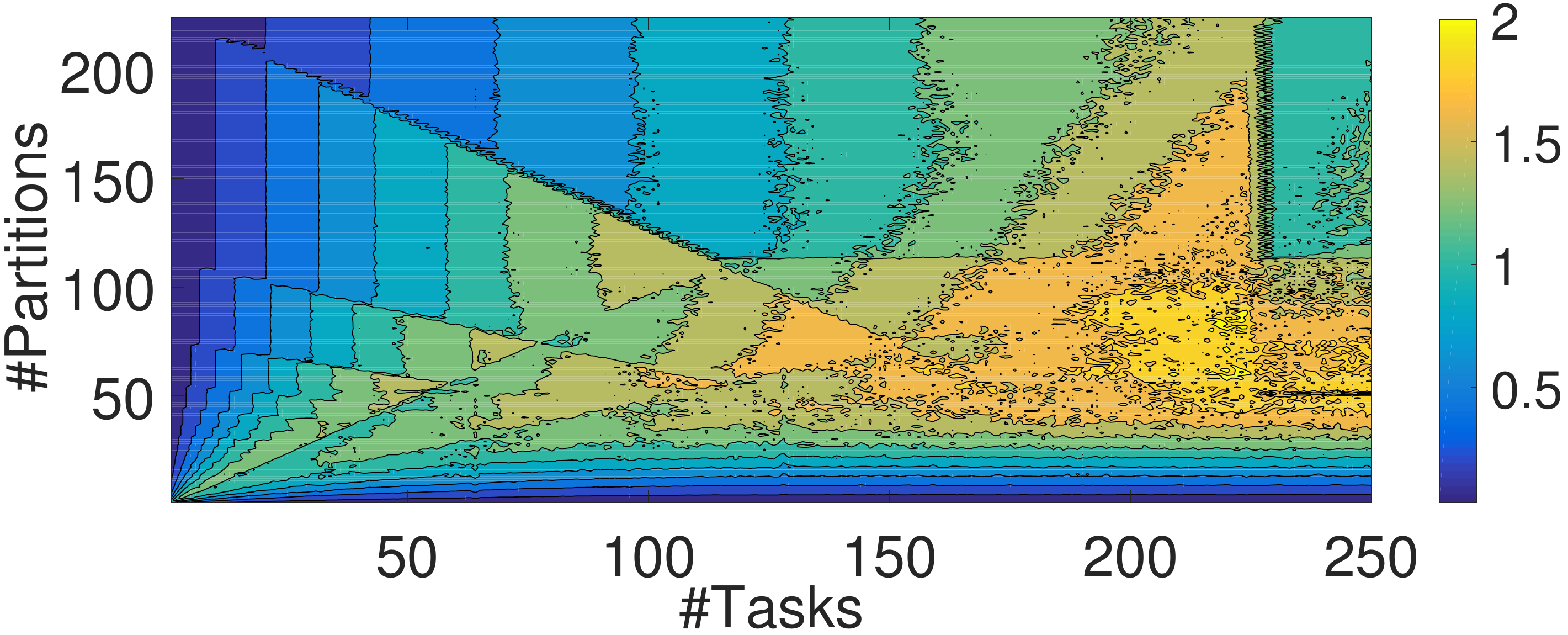}}
\caption{Heatmaps show the resultant speedup (over single-stream) of \texttt{binomial} and \texttt{prefixsum} under different
stream configurations. The \textit{\#partitions} and \textit{\#tasks}  have a significant impact on the resultant performance, and the sweet spots
are sparse and vary across programs. }
\label{fig_motivation_overall}
\end{figure*}

\subsection{Problem Scope}
Our work aims to improve the performance of a data parallel application by exploiting spatial and temporal sharing of heterogeneous
streams. We do so by determining at runtime how many partitions should be used to group the cores (\emph{\#partitions}) and how many data
parallel tasks (\emph{\#tasks}) should be used to run the application. Our current implementation is applicable to XeonPhi and GPUs by
using different runtime back-ends (\textsc{hStream} for XeonPhi, and CUDA or OpenCL for GPUs).

\vspace{2mm} \noindent \textbf{Code Example.} Figure~\ref{fig:example_code} gives a simplified code example written with Intel's \hStreams
APIs that can run on the XeonPhi many-core. At line 2 we initialize the stream execution by setting the number of partitions and
tasks/streams per partition. This initialization process essentially creates multiple processor domains and determines how many logical
streams can run on a partition. In the \emph{for} loop (lines 7-14) we enqueue the communication and computation tasks to a number of
streams identified by the \texttt{stream\_id} variable. In this way, communication and computation of different streams can be overlapped
during execution (temporal sharing); and streams on different processor domains (or partitions) can run concurrently (spatial sharing). Our
predictive model determines the \textit{\#partitions} and the \textit{\#tasks} before invoking the \hStreams initialization routine,
\texttt{hStreams\_app\_init()}. 

\begin{figure}
  \centering
  \includegraphics[width=0.5\textwidth]{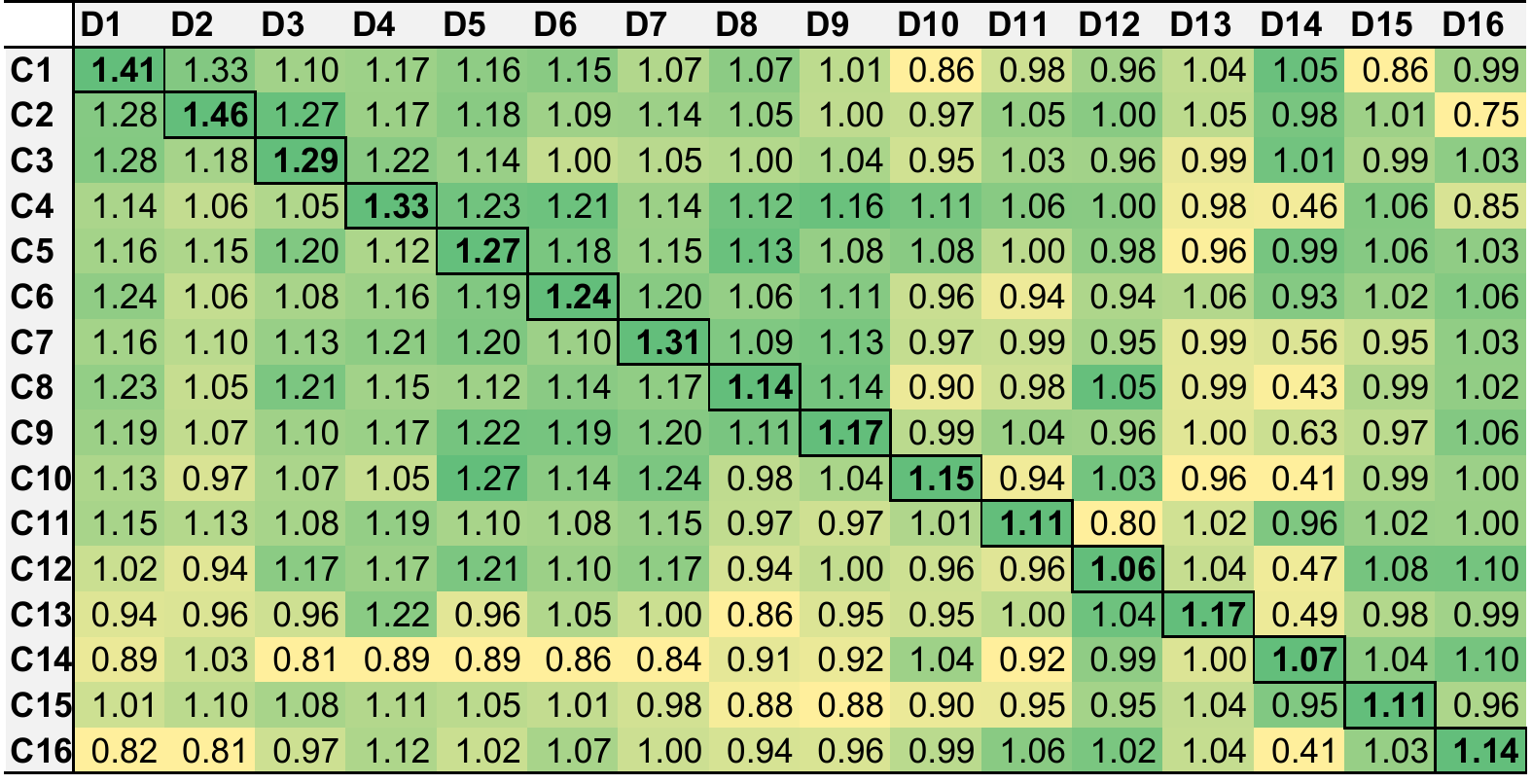}\\
  \caption{Color table showing the speedups of best-performing configurations across inputs for \texttt{dct}. Each cell shows
  the performance for one of the 16 best-performing configurations, $Cn$, on a given input, $Dn$.
  The best configuration varies across inputs and a good configuration on one input
  can give poor performance on another dataset.}\label{fig:acrossdatasets}
  \vspace{-3mm}
\end{figure}

\subsection{Motivating Examples} \label{subsec:motivate:example}

Consider Figure~\ref {fig_motivation_overall} which shows the resultant performance improvement given by multi-stream parallelism over the
single-stream version of the code for two applications on a 57-core Intel XeonPhi system. We use two streamed programs from prior
work~\cite{DBLP:conf/npc/LiFTCY16}: \texttt{binomial} computes the price evolution over a given period and \texttt{prefixSum} calculates
the prefix sum for a sequence of numbers.


It is observed from this example that not all
multi-stream configurations give improved performance. As can be seen from the diagrams, the search space of multi-stream configurations is
huge but good configurations are sparse. The performance varies significantly over stream configurations (\textit{\#partitions},
\textit{\#tasks}). The optimal \textit{\#tasks} for \texttt{binomial} ranges from 1 to 30, and the best \textit{\#partitions} is between  1
and 40. In contrast to \texttt{binomial}, \texttt{prefixsum} benefits from fine-grained parallelism when using a larger \textit{\#tasks}
(220 to 224) and \textit{\#partitions} (60 to 80). However, the stream configurations that are effective for \texttt{prefixsum} give no
speedup over the single-stream version for \texttt{binomial}.

Now consider Figure~\ref{fig:acrossdatasets} that shows the speedups of \texttt{dct} under 16 multi-stream configurations over the
single-stream version, where each configuration is found to give the best-performance for one of the 16 inputs. In the color table, each
cell shows the performance of a stream configuration ($C1, ..., C16$) on a specific input dataset ($D1, ..., D16$); and the values along
the diagonal line represent the best-available performance (found through profiling) for an input. As can be seen from the figure, the best
stream configuration can vary across inputs for the same benchmark. For example, while $C4$ gives a speedup of 1.33x over the baseline for
dataset $D4$, it delivers a poor performance for dataset $D14$ by doubling the execution time over the single-stream version. This diagram
also suggests that no single configuration can give improved performance for all inputs.


\cparagraph{Lesson Learned.} These two examples show that choosing the stream configuration has a great impact on performance and the best
configuration must be determined on a per-program and per-dataset basis. Later, we will show that this observation is not unique to XeonPhi but also holds for GPUs. Attempting to find the optimal configuration through means of an exhaustive search would be ineffective, and
the overhead involved would be far bigger than the potential benefits. Online search algorithms, while can speed up the search process, the
overhead can still outweigh the benefit. For example, when applying simulated annealing to \texttt{binomial}, the best-found configuration
only reaches 84\% of the best-available performance after 310,728 iterations\footnote{In Section~\ref{sec:overall}, we show that our
approach achieves 93\% of the best-available performance for \texttt{binomial} on XeonPhi.}. Classical hand-written heuristics are not
ideal either, as they are not only complex to develop, but are likely to fail due to the variety of programs and the ever-changing hardware
architecture. An alternate approach, and the one we chose to use, is to use machine learning to automatically construct a performance model
to estimate the benefit of any candidate configuration, providing minimal runtime overhead for searching for a good configuration, and
having little development cost when targeting new architectures.

\subsection{Overview of Our Approach}
\begin{figure}
  \centering
  \includegraphics[width=0.5\textwidth]{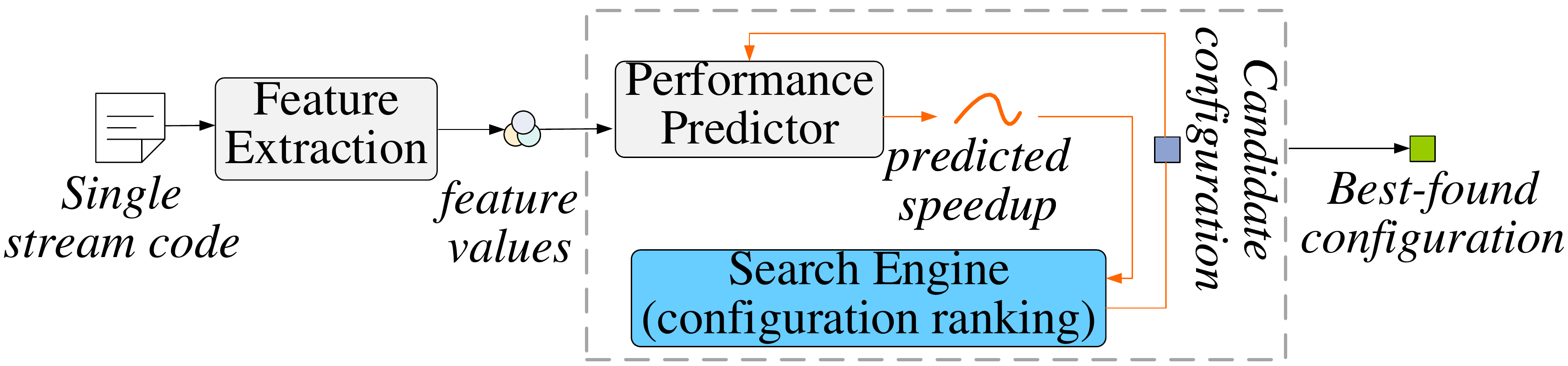}\\
  \caption{Our machine learning based  performance model (trained \emph{offline})
  predicts the speedup based on the extracted feature values of the code and a given stream configuration.
  We use the predictions to quickly rank candidate configurations at runtime to choose the one with the best predicted performance.}\label{fig:workflow}
\end{figure}

Our library-based approach, depicted in Figure~\ref{fig:workflow}, is completely automated. To determine the best streaming configuration,
our approach follows a number of steps described as follows. We use a set of information or \emph{features} to capture the characteristics
of the program. We develop a LLVM~\cite{Lattner:2004:LCF:977395.977673} compiler pass to extract static code features at compile time, and
a low-overhead profiling pass to collect runtime information at execution time (i.e., during the first few loop iterations). Because
profiling also contributes to the final program output, no computation cycle is wasted. At runtime, we search for a good configuration
through an offline trained performance model to estimate the resulting performances for all candidate configurations. The performance model
takes in the feature values, a given configuration of resource partition and task granularity  and estimates the potential speedup for the
given configuration over the single-stream version. The overhead of runtime feature collection and search is a few milliseconds, which is
included in all our experimental results. Since our training process can be performed automatically, we can easily target our performance
model for different architectures.

\section{Performance Modeling} \label{sec_mlstream_modeling}
At the core of our approach is a machine learned performance model built upon the Multi-layer Perceptron (\MLP) artificial neural network
(\ANN). Our prototype is implemented using the Python scikit-learn machine learning package~\cite{scikitlearn}. It is to note that our
prior work~\cite{ipdps18} uses a Support Vector Machine (\SVM) based classifier. However, such an approach can only make predictions on a
limited set of configurations seen at the training time. Unlike a classification-based approach, the new approach presented in this article
is a \emph{regression-based} model which can make predictions on any stream configuration. This new approach thus has a better
generalization ability for various heterogeneous architectures.
We have also evaluated a number of alternative modeling techniques, including \MLP, \SVM, and decision trees. We
chose \MLP because it gives the best performance and has modest training overhead (see Section~\ref{sec_compare_learning_techniques}).

Our performance model takes as input the feature values and a given configuration (e.g., \textit{\#partitions} and
\textit{\#tasks} for XeonPhi and \textit{\#tasks} for GPUs). It predicts the speedup for the given configuration.
Building and using such a model follows a 3-step process for supervised learning: (i) generate training data (ii) train
a performance model (iii) use the performance model, described as follows.

\begin{figure}
  \centering
  \includegraphics[width=0.5\textwidth]{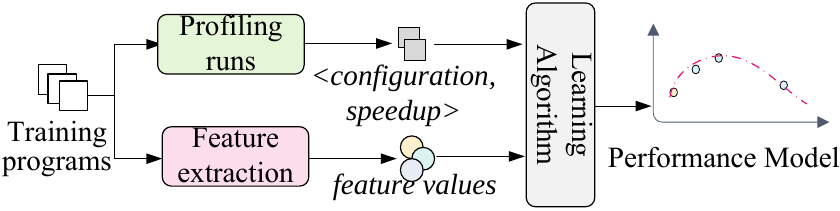}\\
  \caption{The training process of our performance model.}\label{fig:training}
\end{figure}

\subsection{Training the Performance Model\label{sec:training}}
Our method for model training is shown in Figure~\ref{fig:training}. To learn a regression model, we first need to profile the execution
time (in order to calculate the speedup over the single-stream version) of all candidate configurations for each training program, and
extract the feature values from the program. We then use the feature values, configuration settings and speedups to train a model.

\subsubsection{Generating Training Data} To generate training data, we apply \emph{cross-validation} to 39 benchmarks, i.e., by excluding
the testing benchmarks from the training dataset (see also Section~\ref{sec:modeleval}). We execute each training program and benchmark a
number of times until the gap of the upper and lower confidence bounds is smaller than 5\% under a 95\% confidence interval setting. We
then calculate the average speedup for a given stream configuration over the single-stream version.  We exhaustively execute each training
program across a wide range of stream configurations, and record the performance of each. Next, we calculate the speedup for each
configuration, program and dataset. Finally, we extract the values of our selected set of features from each program and dataset. We stress
that the trained model can be applied to stream configurations that are not seen in the training phase.

\subsubsection{Profiling Configurations}
During the training phase, we exhaustively execute each training program across a set of streamed configurations. On XeonPhi, we profile
each training program using the \emph{\#partitions} ranging from 1 to 224 (the maximum number of physical threads on XeonPhi) and the
\emph{\#tasks} ranging from 1 to 256~\footnote{We chose these values because configuration settings beyond these values give a poor
performance during our initial evaluation.}.
On GPUs, we cannot configure the number of partitions currently, we set the \emph{\#partitions} to the same as \emph{\#tasks} to be consistent with XenPhi.
On this platform, we also set the \emph{\#tasks} to be range between $2^0$ and $2^{10}$, which is big enough to
include the optimal values according to our experiments. Note that these parameter ranges can be configured by the user.

\subsubsection{Building The Model}
Each evaluated configuration is appended to the feature value vector of a training program to form a model input. The model inputs and the
corresponding speedups (i.e., ground truths) for all training programs are passed to a learning algorithm. The algorithm finds a
correlation between the input vector and the desired prediction. The output of our learning algorithm is an \MLP model where the weights of
the model are determined from the training data. Model parameter tuning is performed on the training dataset for each targeting hardware
architecture, using cross-validation (see also Section~\ref{sec:ptuning}). In our case, the overall training process for all the 39 training
programs (which is dominated by training data generation) takes less than a week on a single machine. Since training is performed only once
``at the factory'', this is a \emph{one-off} cost.


\subsection{Features} \label{sec_mlstream_modeling_features}

Our performance models are based exclusively on code and dynamic features of the target programs. Code features are extracted from the
program source code, and dynamic features are collected using hardware performance counters during the initial profiling run of the target
application.  We restrict us in using hardware performance counters that are commonly available on modern processors such as the data cache
misses to ensure that our approach can be applied to a wide range of many-core architectures.

We considered 38 candidate raw features in this work. Some features were chosen from our intuition based on factors that can affect the
performance such as \texttt{dts} (host-device data transfer size) and \texttt{\#xfer\_mem}, while other features were chosen
based on previous work~\cite{fursin2008milepost,DBLP:journals/taco/WangGO14}.

\subsubsection{Feature Selection} \label{sec_feature_selection}
To build an accurate model through supervised learning, the training sample size typically needs to be at least one order of magnitude
greater than the number of features. In this work, we start from 311 training  samples and 38 raw features, so we would like to reduce the
number of features in use. Our process for feature selection is fully automatic, described as follows.

We first combine several raw features to form a set of combined normalized features, which are able to carry more information than the
individual parts. For example, instead of reporting raw branch hit and miss counts, we use the branch miss rate. Next, we removed raw
features that carried similar information which is already captured by chosen features. To find which features are closely correlated, we
constructed a correlation coefficient matrix using the Pearson correlation coefficient~\cite{citeulike:1449316822}. The closer a coefficient between
two features is to +/-1, the stronger the correlation between the two input features. We removed any feature which had a correlation
coefficient (taking the absolute value) greater than 0.7. Similar features include the number of executed instructions and the number of
E-stage cycles that were successfully completed.

Our feature selection process reduces the number of features to 10 for XeonPhi (see Table~\ref{tbl_features}) and 10 for the
NVIDIA Titan 1080Ti GPU (see Table~\ref{tbl_gpu_features}), where some features are shared. Since our approach for feature selection is automatic, the approach can be applied to other
sets of candidate features. It is to note that feature selection is also performed using cross-validation (see also
Section~\ref{sec:compa}).

\begin{table}[!t]
\scriptsize
\caption{Chosen features for XeonPhi performance model}
\vspace{-5mm}
\begin{center}
\begin{tabular}{lp{0.65\columnwidth}}
\toprule
\textbf{Feature} & \textbf{Description} \\
\midrule
	\rowcolor{Gray}  loop nest & at which level the outermost parallelizable loop lies on\\
loop count & \# of the parallel loop iterations \\
\rowcolor{Gray}  \#xfer\_mem & \# of host-device transfer API calls\\
dts & total host-device transfer size\\
\rowcolor{Gray}  redundant transfer size & host-device transfer size among overlapping tasks\\
max blocks & the maximum number of tasks of the application\\
\rowcolor{Gray}  min task unit &  the minimum task granularity for a partition\\
\# instructions & the total number of instructions of the kernel\\
\rowcolor{Gray}  branch miss & branch miss rate\\
L1 DCR & L1 Data cache miss rate\\
\bottomrule
\end{tabular}
\end{center}
\label{tbl_features}
\vspace{-3mm}
\end{table}

\begin{table}[!t]
\scriptsize
\caption{Chosen features for GPU programs}
\vspace{-3mm}
\begin{center}
\begin{tabular}{lp{0.55\columnwidth}}
\toprule
\textbf{Feature} & \textbf{Description} \\
\midrule
\rowcolor{Gray} Access type 1 & \# array access, whose fastest varying index is an affine function of the block id\\
Access type 2& \#array accesses, whose second or higher dimensional index is an affine function of the block id\\
\rowcolor{Gray}  \#xfer\_mem & \# of host-device transfer API calls\\
host to device transfer size & total host to device transfer size\\
\rowcolor{Gray}  device to host transfer size & total device to host transfer size \\
redundant transfer size & host-device transfer size among overlapping tasks\\
\rowcolor{Gray}  max blocks & the maximum number of tasks\\
\# instructions & the total number of instructions of the kernel\\
\rowcolor{Gray}  divergent branches& \# divergent branches\\
L2 read miss rate & L2 cache read miss rate\\
\bottomrule
\end{tabular}
\end{center}
\label{tbl_gpu_features}
\vspace{-3mm}
\end{table}

\subsubsection{Feature Standardization}\label{sec:fstandard}
Supervised learning typically requires the feature values to lie in a certain range. Therefore, we scaled the value for each of our
features between the range of 0 and 1. We record the maximum and minimum value of each feature found at the training phase, and use these
values to scale features extracted from a new application after deployment. We truncate a value during deployment if the value is outside
the minimum/maximum value range seen during training.  It is to note that we also use the same approach to normalize the model predictions
(speedups) to the range of 0 and 1. In this work, we choose \textit{Z-score} to standardize the training data, and
the details of quantifying the impact of feature engineering methods can be found
in Section~\ref{sec:fengineering}.

\begin{figure}[t!]
  \centering
    \subfloat[XeonPhi]{
  \includegraphics[width=0.42\textwidth]{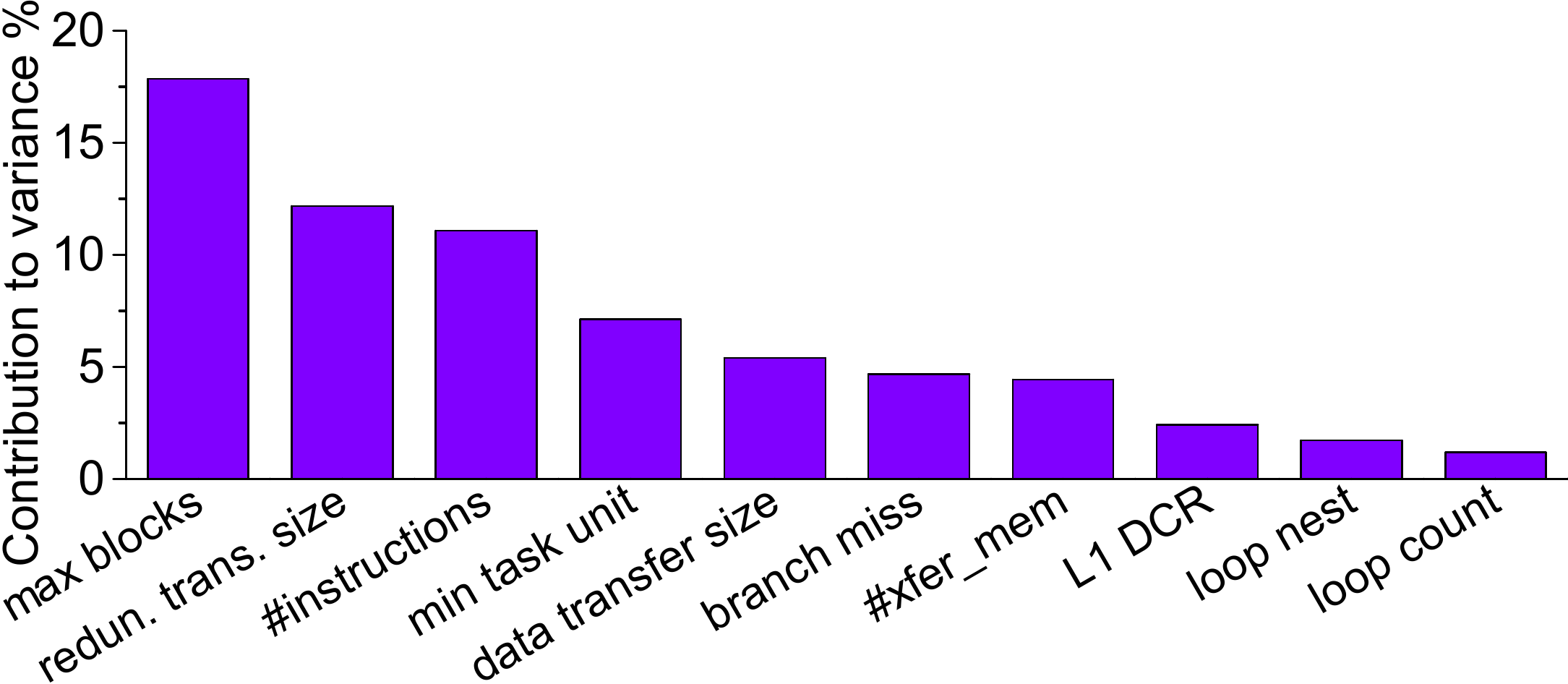}
    }\\
    \vspace{-3mm}
    \subfloat[NVIDIA GPU]{
  \includegraphics[width=0.42\textwidth]{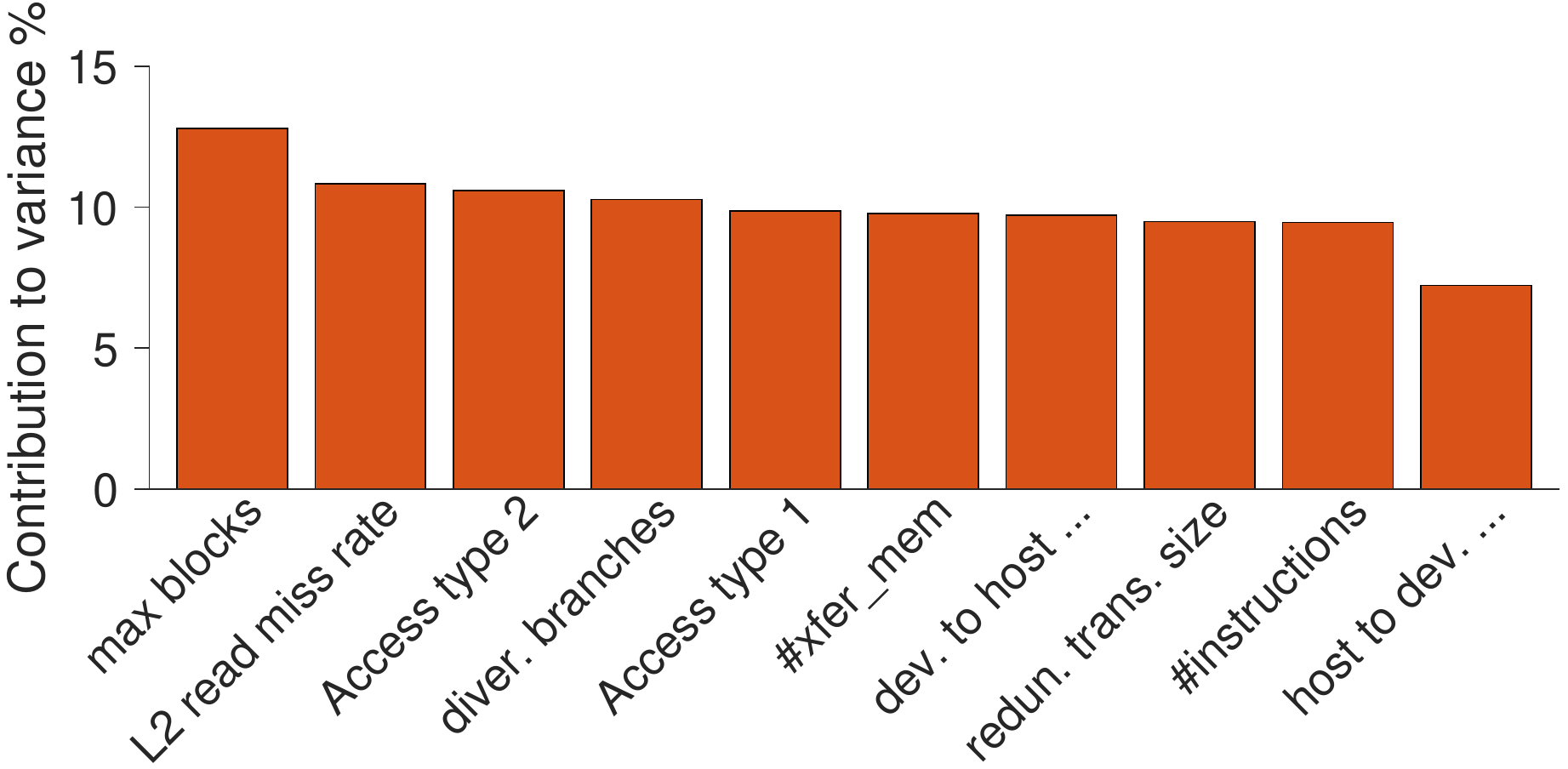} \label{fig:pca-gpu}
    }
	\caption{Feature importance on (a) XeonPhi and (b) NVIDIA GPU.}\label{fig:pca}
  \vspace{-3mm}
\end{figure}

\subsubsection{Feature Importance}
To understand the usefulness\footnote{In Section~\ref{sec:fimp}, we give a further breakdown of the impact of individual feature to the
model performance on a per benchmark basis.} of each feature, we apply a factor analysis technique called Varimax
rotation~\cite{manly2004multivariate} to the feature space transformed by the principal component analysis (\PCA).
 This technique
quantifies the contribution of each feature to the overall variance in each of the \PCA dimensions. Intuitively, the more variances a
feature brings to the space, the more useful information the feature carries.

As an example, Figure~\ref{fig:pca} shows the top features chosen for XeonPhi and NVIDIA GPU architectures. For the XeonPhi platform, features that capture
the parallelism degree (e.g. \texttt{max blocks}), host-device communication (e.g. \texttt{redundant transfer size}), and computation (e.g.
\texttt{\#instructions}) are found to be important. Other features such as \texttt{L1 DCR} and \texttt{loop nest} are useful, but are less
important compared to others.
On the NVIDIA GPU platform, we note that the parallelism degree is important, and
the other features are equally useful (Figure~\ref{fig:pca-gpu}).
This figure shows that prediction can accurately draw upon a subset of aggregated feature values.

\subsection{Runtime Deployment}
 Once we have built and trained our performance model as described above, we can use it as a cost function to search for the best stream configuration for any \emph{new}, \emph{unseen} program.
Feature values are extracted from the single-stream version of the code. Static code features (such as \texttt{loop count}) are extracted
from the program source at compile time. Dynamic features (such as \texttt{branch miss}) are extracted by profiling the program without
partitioning for a few loop iterations (which typically translate to several microseconds). After feature collection, we feed the feature
values to the search engine to rank all candidate configurations using the performance model. The top-ranked stream configuration is then
used for the target program. In Section~\ref{sec:host-code}, we provide further details on how the performance model can be integrated with
the host code generation process.

\subsubsection{Adapt to Changing Program Phases}
Our current implementation chooses a configuration for each kernel and does not change the configuration throughout the
kernel execution. Therefore, it can adapt to different behaviors across kernels because predictions are performed on a
per-kernel basis. We found that this strategy is sufficient for many data-parallel kernels targeted in this work.

Our approach can be extended to adapt phase or program behavior changes within a kernel. One way of doing this is to first partition the
input data into groups and then perform configuration selection before launching the kernel that performs on an input data group. To reduce
the prediction and configuration overhead, we can sample periodically to see if the performance counter readings are significantly
different from the ones used for the current prediction to trigger re-configuration.  Dynamic re-configuration of a running kernel will
require extending the underlying runtime (e.g., \hStreams or CUDA) to adjust thread mapping and having hardware support to stop and resume
the execution contexts. We leave this as future work.

\section{OpenMP to Streamed Code Generator} \label{sec_omp2hstream}
\begin{figure}[!t]
\centering
\includegraphics[width=0.5\textwidth]{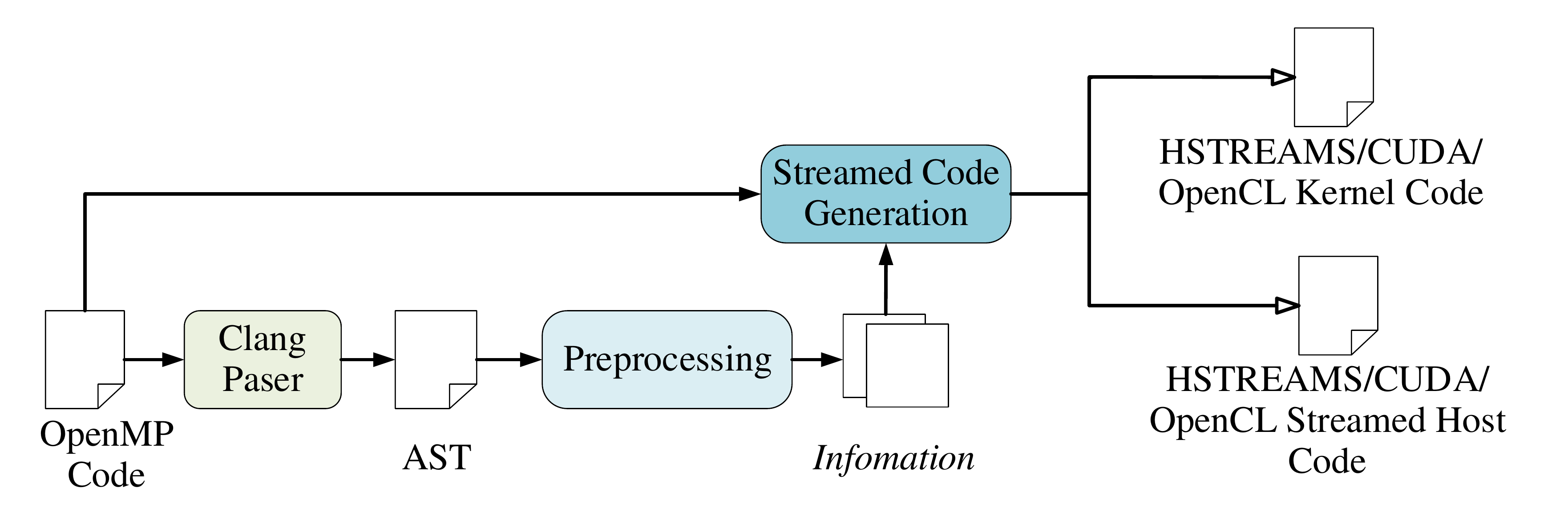}
    \caption{Work flow for translating OpenMP programs to streamed programs using our automatic code generator. \label{fig:omp2hstream}}
\end{figure}

Currently, there are very few publicly available benchmarks for utilizing the streaming capability of heterogeneous many-core
architectures, in particular, XeonPhi. To evaluate our approach on a diverse set of benchmarks, we have developed a compiler-based code
generator, \textsc{autostreamer}, to automatically translate OpenMP programs onto streamed code depending on the target architecture. Our
code generator is open sourced\footnote{Available at: \url{https://github.com/wisdom-moon/autostreamer}.}. Our implementation currently supports converting OpenMP
code to \hStreams, CUDA and OpenCL programs. While we do not claim novelty on this as several works on source-to-source translation from
OpenMP to CUDA\cite{lee2009openmp,grewe2013portable,mikushin2014kernelgen,grosser2016polly} or
OpenCL\cite{DBLP:journals/taco/WangGO14,sotomayor2017automatic} exist, we believe the tool could serve as a useful utility for translating
OpenMP programs to exploit multi-stream performance on heterogeneous many-core architectures.



\subsection{Code Generator Overview}
Figure~\ref{fig:omp2hstream} depicts our source to source code generator for translating OpenMP code to streamed programs. We use LLVM's
Clang front-end to convert OpenMP code into the abstract syntax tree (AST). We then traverse the AST to obtain the information to generate
candidate streamed kernels and host-device management code. The generated kernel and host code make use of exiting programming models for
kernel launching and communication management. We use \hStreams for XeonPhi and CUDA or OpenCL for GPUs.

 Our current implementation
supports the translation of OpenMP parallel loops, i.e., loops annotated with \texttt{omp for} or \texttt{omp for reduction} constructs.
For each parallel loop, we outline the loop body and translate it into an individual kernel function. We then replace the original loop
body with a function call (running on the host CPU) to launch the generated kernel. We also generate management code for streaming context
initialization, data partitioning, data movements between the host and the accelerator, etc.

Our code generator relies on the native host/device compiler to optimize the generated code. We have also compared our automatically
generated code against the manually translated code used in our prior work~\cite{ipdps18} and found that there is little difference in
performance for the set of OpenMP benchmarks used in this work.

%
%

\subsection{Preprocessing}

As an example, Figure~\ref{fig:vecadd} illustrates how an OpenMP parallel loop can be translated into \hStreams
code for XeonPhi. Note that a similar code generation process is implemented for GPUs, using CUDA for NVIDIA GPU
architectures and OpenCL for other GPU platforms.

For each OpenMP parallel loop, we extract information of loop iterations from the loop head. In this work, partitioning is achieved by
splitting the loop iteration space. Furthermore, we collect all the variables needed by the \hStreams kernel. Because \hStreams requires kernel
parameters to be passed as the \texttt{uint64\_t} (lines 1-2 of Figure~\ref{subfig:kernel}), the kernel parameters will be cast into this
type.
The kernel parameters need to be packed into an array (line 21 in Figure~\ref{subfig:host}).
Then the \hStreams library will unpack kernel parameters from the array and pass the parameters to kernel function.

During the preprocessing stage, we also extract the static code feature values of each target parallel loop. The code feature values will be
encoded into the source code during host code generation. It is to note that our approach can be easily applied to existing \hStreams
programs -- by first gathering feature values from an \hStreams kernel, and then storing the extracted information in an auxiliary file or
source code through a compiler front-end pass.

\lstset{}
\begin{figure}
\centering %
\subfloat[OpenMP code.] {
        \noindent\mbox{\parbox{\columnwidth}{%
\begin{lstlisting}[label=subfig:omp]
// An OpenMP C code for vector addition
float * hostOutput = (float *) malloc(inputLength*sizeof(float));
...
#pragma omp parallel for
for(int i=0; i<inputLength; i++)
{
  hostOutput[i] = hostInput1[i] + hostInput2[i];
}
...
\end{lstlisting}%
        }}
  }
\vfil
\subfloat[\hStreams kernel code.] {
        \noindent\mbox{\parbox{\columnwidth}{%
\begin{lstlisting}[label=subfig:kernel]
COINATIVELIBEXPORT void kernel (uint64_t arg0,
       uint64_t arg1, ... uint64_t arg5)
{
  int _start = (int) arg0;
  ...
  float *hostInput2 = (float *) arg5;

  #pragma omp parallel for
  for(int i= _start; i< _end; i++)
    hostOutput[i] = hostInput1[i] + hostInput2[i];
}
\end{lstlisting}%
        }}
  }
\hfil
\subfloat[\hStreams host code.] {
        \noindent\mbox{\parbox{\columnwidth}{%
\begin{lstlisting}[label=subfig:host]
//output buffer
float * hostOutput = (float *) malloc(inputLength*sizeof(float));

//Feature update and prediction
Stream config;

conf_search(&config, &kernel_1_features, kernel_1_profile_runs);
int partitions = config.partitions;
int tasks = config.tasks;

//hStreams Initialization
hStreams_app_init(partitions, 1); .
...
hStreams_app_create_buf((float *)hostInput1, ...);
...

//Work partition
int sub_blocks = inputLength / tasks;
int remain_index = inputLength % tasks;

//Initialize kernel arguments
uint64_t args[6]; args[2] = (uint64_t) inputLength;
...
for (int idx = 0; idx < tasks; idx++) {
   args[0] = (uint64_t) _start;
   _end = _start + sub_blocks;

   if (idx < remain_index)
    _end ++;

   args[1] = (uint64_t) _end;
   hStreams_app_xfer_memory(&hostInput1[_start], &hostInput1[_start], (_end-_start)*sizeof(float), idx % partitions, HSTR_SRC_TO_SINK, NULL);
   hStreams_app_xfer_memory(&hostInput2[_start], ...);

   //Kernel launch
   hStreams_EnqueueCompute(idx % partitions, "kernel_1", 3, 3, args, ...);

  //Read back results
   hStreams_app_xfer_memory(&hostOutput[_start], ...);
   _start = _end;
}
...
//hStreams cleanup code
hStreams_app_fini();
\end{lstlisting}%
        }}
  }
\caption{A running example of translating (a) an OpenMP parallel loop to (b) \hStreams kernel  and (c) host management code. }%
\label{fig:vecadd}%
\end{figure}

\subsection{Kernel Code Generation}
Generating a streamed kernel function is straightforward as much of the OpenMP code can be re-used.
Figure~\ref{subfig:kernel} gives an example of the automatically generated kernel for the OpenMP loop given in
Figure~\ref{subfig:omp} for \hStreams kernels.

For the example given in Figure~\ref{fig:vecadd}, an \hStreams kernel starts with a pre-processor macro \texttt{COINATIVELIBEXPORT} (lines
1-2 in Figure~\ref{subfig:kernel}). The number and the type of the kernel parameters are loop-specific and are automatically determined by
our code generator. Within the generated kernel, all the function parameters are cast from \texttt{uint64\_t} into an appropriate type
before they are used. Note that the OpenMP \texttt{parallel for} pragmas are kept in the generated kernel code per \hStreams requirement
(line 8 in Figure~\ref{subfig:kernel}).

With our code generator, the original outer-most loop iteration space will be partitioned among parallel streams.
The amount of work given to a specific stream is determined by the \texttt{\_start} and \texttt{\_end} variables, which
define which part of the loop iteration space a stream instance will work on. 
A similar kernel code generation approach is implemented for GPUs using CUDA or OpenCL.

\vspace{-2mm}
\subsection{Host Code Generation\label{sec:host-code}}
To generate host code, we replace the original OpenMP parallel loop with a function call to invoke the generated kernel (e.g.,
\texttt{hStreams\_EnqueueCompute} in Figure~\ref{subfig:host})) together with additional code to initialize the host context and to manage
data transfer.

\vspace{-2mm}
\subsubsection{Feature Value Collection}
Static code features, extracted by our code generator, will be encoded as a feature vector of real values. The feature vector will be
passed to our configuration search engine to find the optimal stream configuration at runtime. Dynamic feature values are automatically
collected by running the generated streamed kernel for 5 iterations under the single-stream configuration. As some loop bounds are
dependent on the input, we might be unable to determine certain feature values at compile time. These features are represented as static
symbolic pre-computation of loop bound variables, which will be updated using runtime values at runtime.

\subsubsection{Setting Stream Configurations} To partition tasks among streams, we break the loop iterations into a number of chunks of an equal size of subtask. We then
group the hardware processor cores into partitions, where each partition contains a fixed set of streams. Processor
partitioning and streams creation are achieved by calling the \texttt{hStreams\_app\_init} (line 12 in Figure~\ref{subfig:host}) function for
XeonPhi (and \texttt{cudaStreamCreate} and \texttt{clCreateCommandQueue} for CUDA and OpenCL programs respectively) by passing the stream configuration given
by our search engine. To overlap host device communications, we further split the input/output data arrays to multiple
data blocks (lines
 32-39 in Figure~\ref{subfig:host}) where each task operates on one block at a time while another data block is transferring between
the host and the accelerator. The number of data blocks is determined by the stream configuration chosen at program
runtime. The amount of work per task
and the size of transferred data can be determined
with kernel parameters.
For example, in \emph{for-loop}
at line 24 of Figure~\ref{subfig:host}, we calculate them
with the starting position (\texttt{\_start})
and the block size (\texttt{sub\_block}).
Thereafter, we schedule tasks and transfer the corresponding data blocks onto streams in a round-robin fashion.

\vspace{-2mm}
\subsubsection{Runtime Prediction}
When a streamed (e.g., \hStreams or CUDA) kernel is invoked, the configuration selection engine library will choose a
stream configuration (line 7 in Figure~\ref{subfig:host}) for the kernel. It uses the performance model to rank the
candidate stream configurations and returns the optimal configuration (\emph{\#partitions} and \emph{\#tasks} for the
example shown in Figure~\ref{fig:vecadd}). The returned values are then used to initialize the streamed context (lines
8-9 of Figure~\ref{subfig:host}). The overhead of prediction is negligible (a few milliseconds) and is included
in the results.

\vspace{-2mm}
\subsubsection{Supporting OpenMP Constructs} OpenMP variables may have additional type information specified by directives, including
\texttt{default}, \texttt{share}, \texttt{private}, \texttt{firstprivate}, \texttt{lastprivate}, \texttt{copyin} and
\texttt{threadprivate}. Our generator uses these directives to map data onto the accelerator memory space. Each variable with the
\texttt{share} or \texttt{default} directive will be translated into a global variable shared by all parallel threads. Variables declared
as \texttt{private} and \texttt{threadprivate} are translated such that there is a private copy for each streamed kernel; no memory
transfer between the host and the accelerator is needed. For each variable specified as \texttt{copyin} or \texttt{first private}, we
create a private copy for each streamed kernel but initialize each copy using explicit memory transfers before its first use. Similarly, we
create a private copy of a \texttt{last private} variable and the original variable is updated by a stream that executes the last
iteration.

Our implementation also supports a number of synchronization and thread constructs. Structured blocks identified with
\texttt{master}, and \texttt{single} directives are executed by one thread on the host multi-core. \texttt{barrier} is
implemented by splitting up the parallel loop into smaller tasks to create synchronization points among multiple
streams. \texttt{critical} is implemented by using a mutex lock to restrict the execution of the associated structured
blocks to a single thread at a time. The \texttt{atomic} and \texttt{flush} directives are already supported by
\hStreams, CUDA or OpenCL.

\subsubsection{Host-Accelerator Communication Optimization}
 For each buffer that is used by both the host and the accelerator, we manage two copies: one on the host memory and the
other on the accelerator memory. Our runtime records the status of each variable and checks whether the copy on a device memory space is
valid or not. No memory transfer is needed as long as the copy in the target memory space is valid. We currently use a conservative
approach: if an element of an buffer has been updated, the entire buffer needs to be synchronized before it can be used by threads running
on a different device. We also avoid unnecessary device to host data transfer by tracking the data dependence between the kernel and the
host program. For example, when there are data-dependencies between two kernels but the host does not access this data in between the two
kernels, we directly pass the memory address of the buffer to the later kernel (without moving the data back to the host).

%
%

\section{Experimental Setup} \label{sec_mlstream_setup}
\vspace{-1mm}
\subsection{Hardware, Systems Software and Benchmarks} \label{subsec:benchmarks}
\begin{table}[t!]
  \centering
  \scriptsize
    \caption{Our evaluation platforms}\label{tbl:evalplatform}
  \begin{tabular}{lll}
  \toprule
  & CPU-XeonPhi & CPU-GPU \\
  \midrule
  CPU & 8-core Xeon CPU @ 2.6 GHz & Core i7-8700K CPU @ 3.7 GHz\\
  Accelerator &   Intel Xeon 31SP Phi & NVIDIA GeForce GTX 1080 Ti GPU\\
  \bottomrule
  \end{tabular}

\end{table}

\vspace{-1mm}
 \cparagraph{Platforms.}
We evaluate our approach on two heterogeneous many-core platforms: one is a CPU-XeonPhi platform and the other is a
CPU-GPU platform. Table~\ref{tbl:evalplatform} gives details of our hardware platforms.

\cparagraph{Systems software.} On the CPU-XeonPhi platform, the host CPU and the accelerator are connected through
\texttt{PCIe}. The host runs Redhat Linux v7.0 (with kernel v3.10). The coprocessor runs a
customized uOS (v2.6.38.8). We use Intel's MPSS (v3.6) to communicate between the host and the coprocessor. We use the
Intel ~\hStreams library (v3.6) and Intel ICC ({v16.0.3}) for compilation (with -O3 as the compiler option).
The CPU-GPU platform runs Ubuntu 16.04 (with kernel v4.15).  We use CUDA v10.0 and gcc v5.3 as the host compiler
with option ``-O3".

\cparagraph{Benchmarks.} We use our code generator to translate 37 OpenMP applications from commonly used benchmark
suites into \hStreams and CUDA programs. We have excluded benchmarks where the data transfer cannot be
overlapped with the kernel execution, which do not benefit from streamed parallelization. Table~\ref{tbl_benchmarks}
gives the full list of these benchmarks. Among them, \texttt{convolutionFFT2d} and \texttt{convolutionSeparable} have
algorithm-dependent parameters, which are regarded as different benchmarks in the experiments. This setting gives us a
total of 39 programs. We run the majority of the programs using over 25 different datasets, except for some
applications where we used around 10 datasets because the algorithmic constraints prevent us from using a larger number
of inputs.


\begin{table}[!t]
\scriptsize
\caption{Streamed benchmarks used in our experiments.}
\vspace{-5mm}
\begin{center}
\begin{tabular}{lllll}
\toprule

\textbf{Suite} & \textbf{Name} & \textbf{Acronym} & \textbf{Name} & \textbf{Acronym}\\
\midrule

\rowcolor{Gray}         & convol.Separable  & convsepr1(8) & dotProduct &  dotprod  \\
\rowcolor{Gray}         & convolutionFFT2d  & fftx1y1(4y3) & fwt &  fwt  \\
\rowcolor{Gray}         &  MonteCarlo & montecarlo   & matVecMul  & mvmult  \\
\rowcolor{Gray}         & scalarProd &  scalarprod &  transpose  & transpose \\
\rowcolor{Gray}         \multirow{-5}{0.01\textwidth}{NVIDIA SDK}
                        & vectorAdd &  vecadd  & &\\

\multirow{2}{0.01\textwidth}{AMD SDK}
       & binomial & binomial  & BlackScholes & blackscholes \\
       & dct  & dct & prefixSum & prefix  \\

\rowcolor{Gray}         & bfs & bfs  & histo&  histo  \\
\rowcolor{Gray}         & lbm & lbm & mri-q &  mri-q  \\
\rowcolor{Gray}         & mri-gridding & mri-gridding &  sad & sad \\
\rowcolor{Gray}         \multirow{-3}{0.01\textwidth}{Parboil}
                        & sgemm &  sgemm & spmv  & spmv \\

\multirow{8}{0.01\textwidth}{POLY BENCH}
       & 2mm & 2mm & 3mm & 3mm \\
	& adi & adi & correlation & correlation \\
       & covariance & covariance & deriche & deriche\\
	 & gemm & gemm & gemver & gemver \\
	& gesummv & gesummv  & heat-3d & heat-3d \\
	& jacobi-1d & jacobi-1d & jacobi-2d & jacobi-2d\\
	& mvt & mvt & syr2k & syr2k \\
	& syrk & syrk & & \\

\bottomrule
\end{tabular}
\end{center}
\label{tbl_benchmarks} \vspace{-3mm}
\end{table}

\vspace{-2mm}
\subsection{Competitive Approaches\label{sec:compa}}
We compare our regression-based approach against our preliminary work that employs an \SVM-based classifier to predict
the optimal stream configuration~\cite{ipdps18}. We denote our prior approach as \SVMC.
We also compare our approach against two recent models for predicting the optimal stream configuration on GPUs. As it is currently not
possible to configure the number of processor partitions  on GPUs, the relevant GPU models can only predict the number of tasks.

\cparagraph{\Modela} In ~\cite{citeulike:13920353}, Liu \etal use linear regression models to search for the optimal number of tasks for
GPU programs~\cite{citeulike:13920353}. The approach employs several analytic models, described as follows.

For a task with an input data size of $m$, the transferring time  between the CPU and the accelerator, $T_t$, is
determined as $T_t = \alpha \cdot m + \beta$, and the computation time, $T_c$, is calculated as: $T_c=\eta \cdot
m+\gamma$ where the model coefficients, $\alpha$, $\beta$, $\eta$ and $\gamma$, are determined through empirical
experiments. For a given kernel with $N$ input data elements running using $n$ streams, this approach partitions the
computation into $n$ tasks, where the data size for each task, $m$, is equal to $N$/$n$.
For the programs which kernel dominated,
the total execution
time, $T_{total}$, can be determined by:
$$T_{total} =T_{t} + nT_{c}=\alpha \cdot m+\frac{N\gamma}{m}+N\eta+\beta$$
For the programs which data transfer dominated:
$$T_{total} =\alpha \cdot N+2\frac{N}{m}\beta$$
By calculating the partial differential and second-order partial differential of
$T_{total}$ with respect to $m$, we can obtain the optimal task-granularity as $m= \sqrt{\frac{N\gamma}{\alpha}}$.
Then we can calculate the number of tasks ($n$).

Note that $m=N/2$ is the optimal parameter for programs which data transfer dominated, i.e., the optimal number of tasks is 2. Another
problem of this model is that it does not consider scenarios where communications in different direction (i.e., host to device and device
to host) can overlap with each other. Note that we set the \textit{\#partitions} to be the same as $n$ for XeonPhi.

\cparagraph{\Modelb} The work presented by Werkhoven \etal models the performance of data transfers between the CPU and the
GPU~\cite{citeulike:13920334}. They use the LogGP model to estimate the host-device data transfer time. Specifically, the model estimates
the data transfer time using five parameters: the communication latency ($L$), overhead ($o$), the gap ($g$), the number of processors
($P$), and the \texttt{PCIe} bandwidth ($G$).

Let $B_{hd}$ denotes the amount of data transferred from the host to the device and $B_{dh}$ denotes vice versa, and $T_{kernel}$ donates
the kernel execution time.
For the dominant transfer scenario, the optimal number of tasks(i.e., \emph{\#tasks}), $N_s$,  can be estimated by solving the following
equations:
\begin{small}
\[
B_{dh} * G_{dh} + g*(N_s-1) =
\begin{cases}
\frac{T_{kernel}}{N_s}+\frac{B_{dh}}{N_s} *G_{dh}, & \text{if} B_{dh} > B_{hd}\\
\frac{B_{hd}}{N_s}*G_{hd}+\frac{T_{kernel}}{N_s}, & \text{otherwise}
\end{cases}
\]
\end{small}
This model does not consider the dominant kernel scenario, as it assumes the kernel execution time will increase as the number of streams
increases and can not model the kernel execution time. Here, we use the same equation to calculate the optimal number of tasks. For this
model, we also set the \textit{\#partitions} to be equal to the optimal $N_{s}$ value on XeonPhi.


\subsection{Evaluation Methodology}

\subsubsection{Model Evaluation\label{sec:modeleval}} We use  cross-validation to evaluate our machine learning models. To test the portability of our approach,
we apply \emph{leave-one-out} cross-validation, described as follows. We exclude the target program for predictions
from the training program set, and learn a model using the \emph{remaining} programs. We then apply the learned model
to the testing program. We repeat this process until each benchmark is tested once. This is a standard evaluation
methodology, providing an estimate of the generalization ability of a machine learned model in predicting \emph{unseen}
data. Note that we exclude both \texttt{convolutionFFT2d} and \texttt{convolutionSeparable} from the training set when
one of the two is evaluated, and we make sure all approaches are trained on the same benchmarks for fair comparisons.

\subsubsection{Performance Report} We run each program under a stream configuration multiple times and report the
\emph{geometric mean} of the runtime. Compared to the arithmetic mean, the geometric mean is often considered as a more
suitable metric for reporting program performance, as it can better minimize the impact of
outliers~\cite{ertel1994definition}. To determine how many runs are needed, we calculated the confidence range using a
95\% confidence interval and make sure that  the difference between the upper and lower confidence bounds is smaller
than 5\%.


\section{Experimental Results} \label{sec_mlstream_results}
In this section, we first present the overall performance of our approach on both platforms.
We then compare our approach to that uses fixed stream configurations, two prior analytical models and our previous work.
We futher discuss the benefit sources of the streaming parallelism and the working mechanism of our approach.
At last, we demonstrate the tunning process of our model.

\subsection{Overall Performance}
\label{sec:overall}
\begin{figure*}[!t]
  \centering
    \subfloat[XeonPhi]{
        \includegraphics[width=\textwidth]{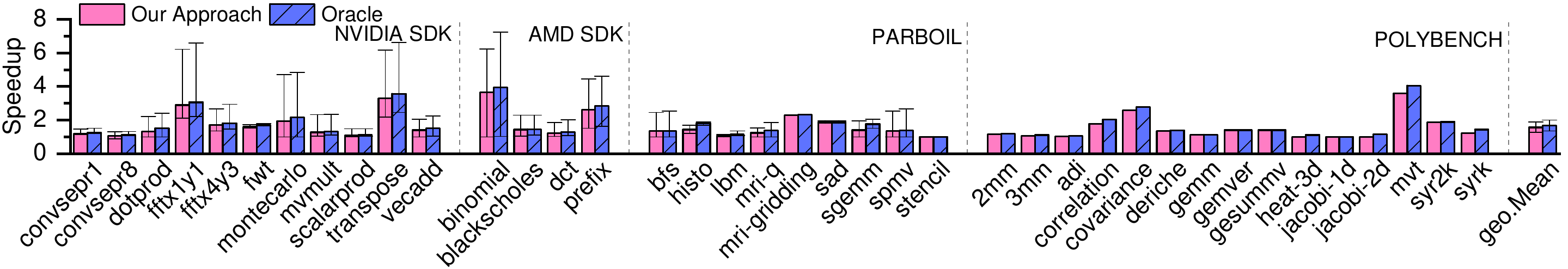}
    }\\
    \subfloat[NVIDIA GPU]{
    \includegraphics[width=\textwidth]{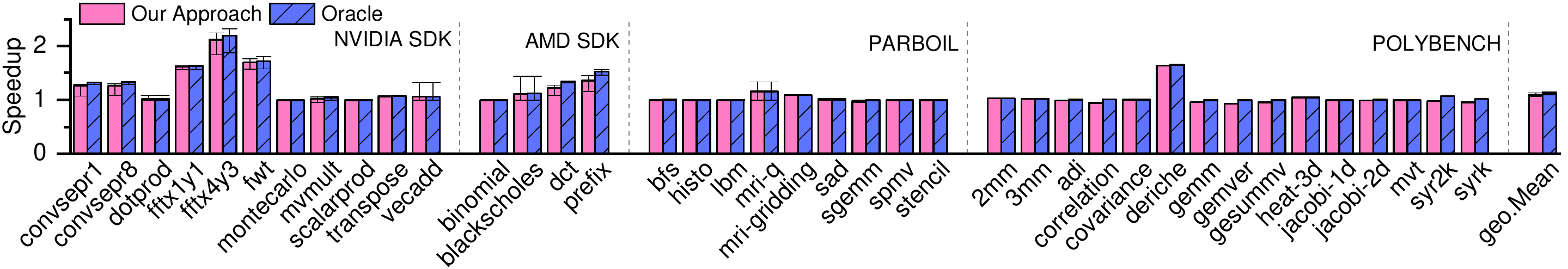}
    }

  \caption{Overall performance of our approach over a single-stream version on XeonPhi (a) and NVIDIA GPU (b). Our approach achieves, on average,
  93.7\% and 97.9\% of the oracle performance on XeonPhi and NVIDIA GPU, respectively.
  The min-max bars show the range of performance achieved across different inputs. }\label{fig_regression_perf}
\end{figure*}


In this experiment, we exhaustively profiled each application with all possible stream configurations and report the
best-found performance as the \emph{Oracle} performance. The Oracle gives an indication of how close our approach is to
a \emph{theoretically perfect} solution. The baseline used to calculate the speedup is running the application using a
single-stream without processor core or task partitioning.

 The overall result is shown in Figure~\ref{fig_regression_perf}. The min-max bar on the diagram
shows the range of speedups per application across all evaluated inputs. Overall, our approach achieves an average
speedup of 1.57$\times$ and 1.1$\times$ over the single-stream configuration on XeonPhi and the GPU respectively. This
translates to 93.7\% and 97.9\% of the Oracle performance on XeonPhi and the GPU respectively.

On XeonPhi, the performance improvement of our approach comes from two factors. First, by predicting the right
processor
 partition size, our approach allows effective overlapping of the host-device
communication and computation.
Second, by matching task parallelism to the number of available processor cores, our
approach can reduce the overhead of thread management, compared to the single-stream execution.
When the host-device
communication time dominates the streaming process, the performance improvement mainly comes from
computation-communication overlapping and the speedup from streaming is consistently less than 2$\times$. When the
kernel execution time dominates the stream process, the application can benefit from the overhead reduction of thread
management. In this case, the speedup can be as large as 5$\times$. We provide a further discussion on this later in
Section~\ref{sec:hsc}.


On the GPU, we can exploit bidirectional data transfer between the host and the device by using pined memory which is not supported by \hStreams.
The support of bidirectional data transfer allows us to obtain further performance gains by overlapping host-device data transfer and computation.
The theoretically up-bound speedup on the GPU platform is 3$\times$, when data transfer is perfectly overlapped with computation.
The representative sample is \texttt{fftx4y3} with the larges dataset, the data transfer time in the two directions is the same, and the kernel execution time is 1.5 times of the data transfer time.
The oracle speedup is 2.3$\times$, and our approach achieves a speedup of 2.2 $\times$.
 On the other hand, because the current GPU implementation does not support processor
core partition, the kernel execution time benefits less from using multiple streams.
Programs which the kernel execution time dominated have no speedup using multiple streams, such as \texttt{bfs}, \texttt{MonteCarlo}.

\subsection{Comparison to Fixed Stream Configurations}
\begin{figure*}[!t]
  \centering
    \subfloat[XeonPhi]{
  \includegraphics[width=\textwidth]{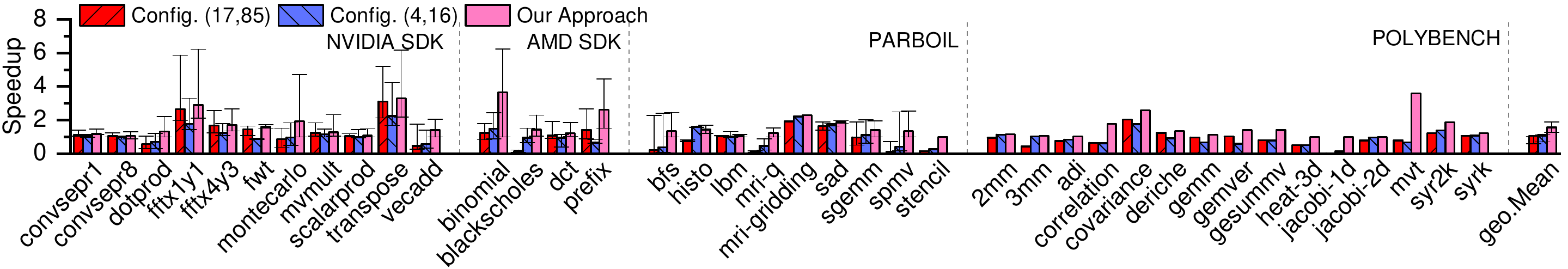}
  \label{fig_fixed_config_a}
    }\\
    \subfloat[NVIDIA GPU]{
  \includegraphics[width=\textwidth]{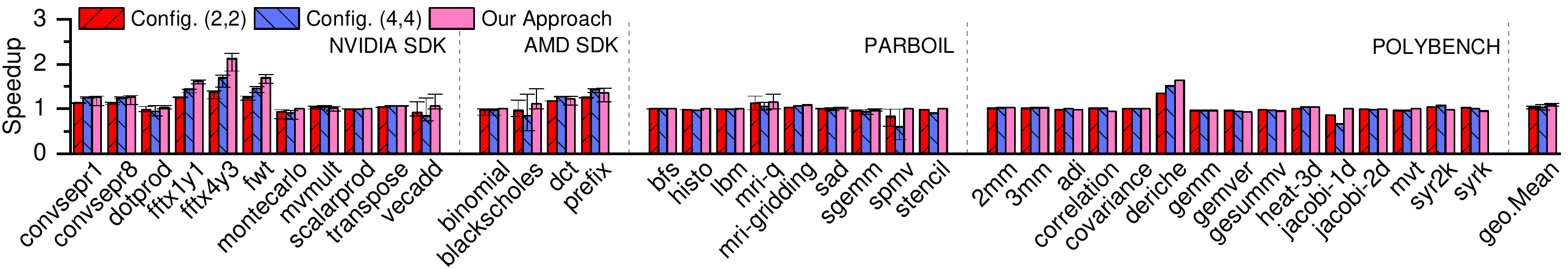}
  \label{fig_fixed_config_b}
    }
  \vspace{-2mm}
  \caption{Comparing the performance with two fixed configurations on XeonPhi (a) and NVIDIA GPU (b): config. $(4,16)$ of 4 partitions and 4 tasks per partition, config. $(17,85)$ of 17 partitions and 5 tasks per partition, config. $(2,2)$ of 2 partitions and 1 tasks per partition, and config. $(4,4)$ of 4 partitions and 1 tasks per partition.}\label{fig_fixed_config}
\end{figure*}

Our approach predicts from a wide range of stream configurations, which configuration is likely to give the best
performance for a given program and dataset. A natural question to ask is that: \textit{is there a fixed stream configuration that
gives reasonable good performance across benchmarks and datasets?} To answer this question, we compare our predictive
modeling based approach to two specific configurations on each of our evaluation platforms. Our justification for
why we selecting the fixed configurations are described as follows. On XeonPhi, our initial results in
Section~\ref{sec_mlstream_motivation} indicate that using the stream configuration of $(4,16)$, i.e. partitioning the
cores to 4 groups and running 4 tasks on each partition (16 tasks in total), gives good performance. The statistics
obtained from the training data suggest that the configuration of $(17,85)$ give the best average performance across
training samples. On the GPU, several programs support a maximum of 4 tasks. Thus we select the two configurations
$(2,2)$ and $(4,4)$. The results are shown in Figure~\ref{fig_fixed_config}.

\begin{figure}[!t]
  \centering
    \subfloat[XeonPhi]{
  \includegraphics[width=0.5\textwidth]{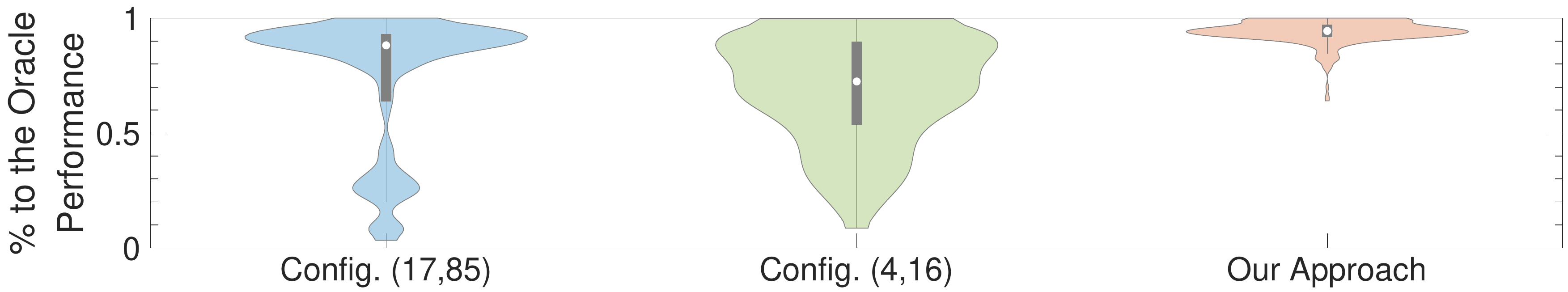}
  \label{fig_upper_bound_a}
    }\\
    \subfloat[NVIDIA GPU]{
  \includegraphics[width=0.5\textwidth]{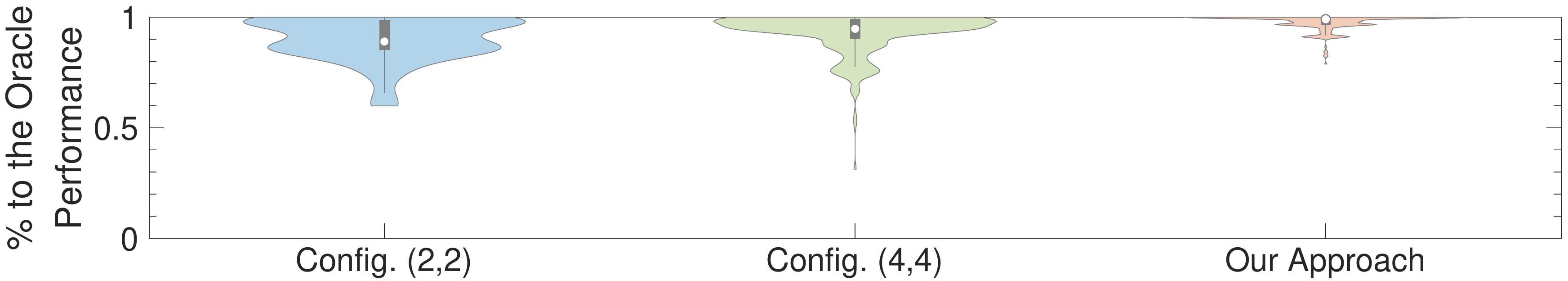}
  \label{fig_upper_bound_b}
    }

  \caption{Violin plot showing the distribution of speedups per scheme across benchmarks and datasets on XeonPhi (a) and GPU (b).
The shape of the violin corresponds to the speedup distribution to the oracle performance. The thick black line shows
where 50\% of the data lies.}\label{fig_upper_bound} \vspace{-4mm}
\end{figure}

\subsubsection{XeonPhi} On XeonPhi, we observe improved performance for several benchmarks such as \texttt{mri-gridding},
\texttt{transpose}, \texttt{sad}, under both configurations, but slower performance for \texttt{dotprod},
\texttt{vecadd}, \texttt{blackscholes}, \texttt{lbm}, and \texttt{mir-q} (Figure~\ref{fig_fixed_config_a}). For \texttt{prefix}, configuration $(17,85)$
delivers improved performance while configuration $(4,16)$ leads to slowdown performance. Overall, none of the two
fixed configurations give an improved performance on average. On average, our approach outperforms the two fixed
configurations by a factor of 1.4, and delivers consistently improved performance across benchmarks and datasets.

The violin plot in Figure~\ref{fig_upper_bound_a} shows how far is each of the three schemes to the Oracle performance
across benchmarks and datasets. Our approach not only delivers the closest performance to the Oracle, but also has the
largest number of samples whose performance is next to the Oracle. By contrast, the performance given by the fixed
configurations for many samples is farther from the Oracle performance.

\subsubsection{GPU}
On the GPU, in most cases, the performance of configuration $(2,2)$ is moderate, not great, but not much worse than single-version, leading
to an average speedup 1.03$\times$ (Figure~\ref{fig_fixed_config_b}). By contrast, although configuration $(4,4)$ performs poorly on two
programs, it delivers a slightly larger averaged speedup of 1.04$\times$. By choosing the stream configuration on a  per-program basis, our
approach outperforms the two fixed configurations, achieving an averaged speedup 1.10$\times$. On only four programs, our approach delivers
slightly worse performance with a small margin.

The violin plot in Figure~\ref{fig_upper_bound_b} also confirms the strengths of our approach by presenting the
distribution of performance improvement. The results on the diagram are normalized to the Oracle (best-available)
performance. For most of the programs, the two fixed configurations deliver 80\% to 100\% to the Oracle performance.
However, configuration $(4,4)$ can lead to rather poor performance (less than 40\% to the best available performance)
on some programs. Compared to the fixed configurations, the performance distribution given by our approach is
centralized on a range between 90\% to 100\%, where most programs are within this percentile range. Furthermore, compared
to the fixed configurations, our approach has a fewer number of performance outliers, which have less serious
performance slowdown. Therefore, our approach delivers consistently better performance compared with the fixed
configurations.

\vspace{-2mm}
\subsubsection{Summary}
This experiment confirms that a fixed configuration fails to deliver improved performance across applications and
datasets, and selecting a right stream configuration on a per program, per dataset basis is thus required.


\vspace{-2mm}
\subsection{Comparison to Analytical Models\label{sec:alt}}

\begin{figure*}[!t]
  \centering
    \subfloat[XeonPhi]{
  \includegraphics[width=\textwidth]{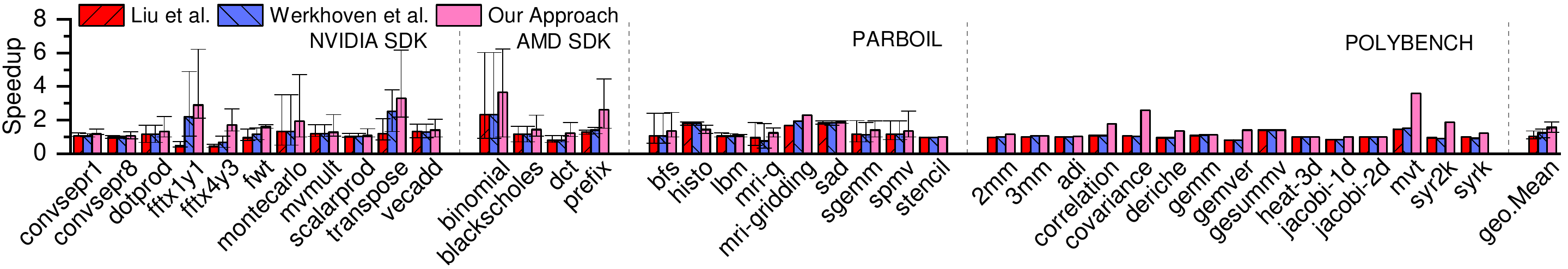}
  \label{fig_hs_model_a}
    }\\
    \subfloat[NVIDIA GPU]{
  \includegraphics[width=\textwidth]{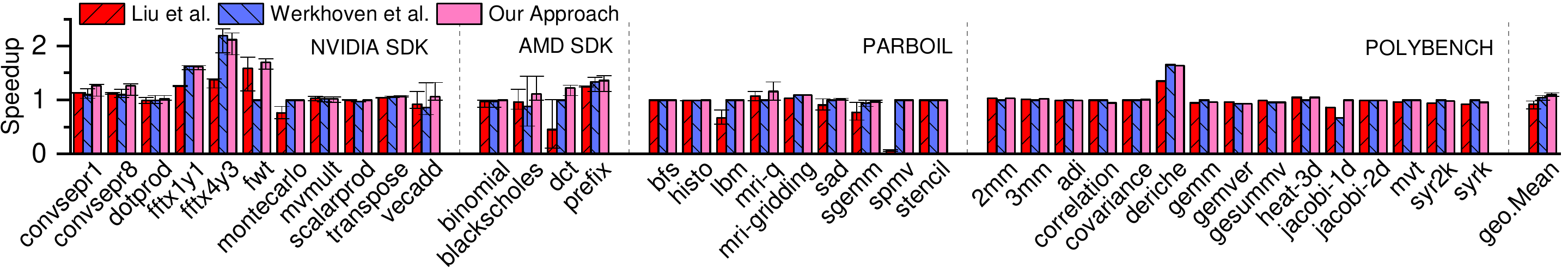}
  \label{fig_hs_model_b}
    }
  \caption{Comparing against \Modela and \Modelb on XeonPhi (a) and NVIDIA GPU (b).
  }\label{fig_hs_model}
\end{figure*}

\begin{figure}[!t]
  \centering
    \subfloat[XeonPhi]{
  \includegraphics[width=0.5\textwidth]{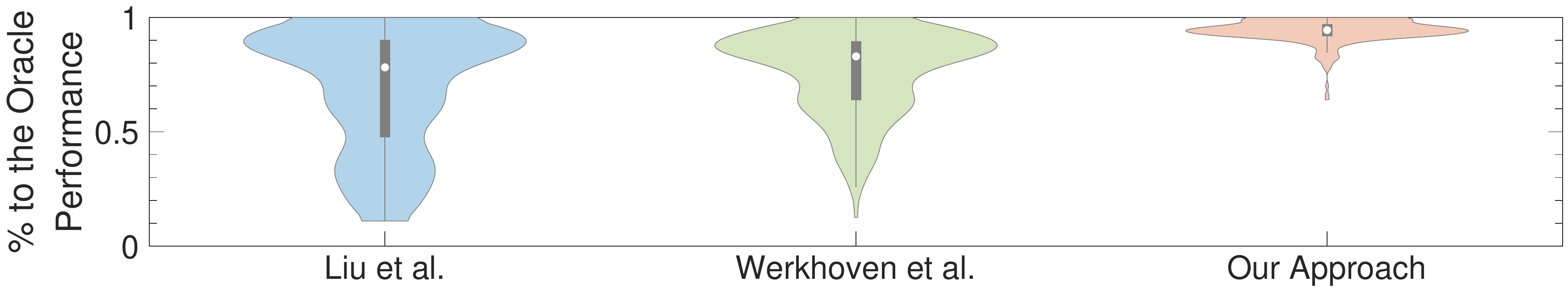}
  \label{fig_vl_model_a}
    }\\
    \subfloat[NVIDIA GPU]{
  \includegraphics[width=0.5\textwidth]{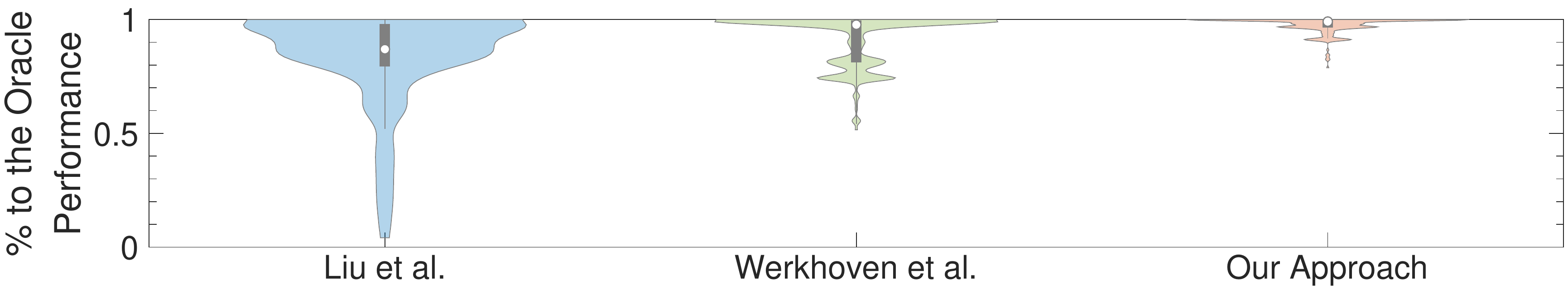}
  \label{fig_vl_model_b}
    }
  \caption{Violin plots showing the distribution of speedups across benchmarks and datasets on XeonPhi (a) and GPU (b).}\label{fig_vl_model}
\end{figure}

In this experiment, we compare our approach to the two recent analytical models described in Section~\ref{sec:compa}.
The results are shown in Figures~\ref{fig_hs_model} and~\ref{fig_vl_model}. On XeonPhi, both competitive models prefer
using $2$ tasks across benchmarks and datasets. This is because that many programs are kernel dominated, the analytical models simply assume that task
partition has no effect on kernel's performance, and do not consider the thread management overhead. On the GPU, the
model proposed by \Modela tends to use $2$ tasks across benchmarks and datasets. This is due to the fact that most programs are data transfer dominated and this
model ignores the overlap of the bidirectional data transfers between the host and the device.

\cparagraph{XeonPhi.} Figure~\ref{fig_hs_model_a} demonstrates that our approach gives better performance for nearly all
programs on XeonPhi. For the remaining handful programs, all three approaches  deliver comparable performance. Compared
to the results Figure~\ref{fig_fixed_config}, we can find the performance of the analytical models is similar to fixed
stream configurations. This is because the performance of the seven programs, such as \texttt{binomial}, changes
dramatically with different stream configurations (see also Figure~\ref{fig_motivation_overall}). The performance of
the remaining programs is not sensitive to the variation of stream configurations. From Figure~\ref{fig_vl_model_a}, we
can further see that \Modela and \Modelb deliver a speedup within a range on 20\% to 80\%, while the performance of our
approach is centralized on a range between 80\% to 100\%. Thus, our approach delivers consistently better performance
compared with the alternative models.

\cparagraph{GPU.} Figure~\ref{fig_hs_model_b} shows that our approach delivers better performance for around 75\% of
the programs on the GPU.  Since \Modelb and \Modela are manually tuned for the GPUs, they give better performance on
some benchmarks over our approach. However, our approach has the advantages of being automatically learned from
training data, with little expert involvement. The performance of our approach can be further improved by using more
training examples to better cover the program space. Figure~\ref{fig_vl_model_b} shows that \Modela and \Modelb
delivers a speedup within a range of 5\% to 80\%, and 70\% to 100\%, respectively. By contrast, the performance of our
approach is centralized within a range between 90\% to 100\% for more programs. Therefore, overall, our approach
delivers better average performance compared with the alternative models.

\subsection{Comparison to Classification-based Approach}
\begin{figure*}[!t]
  \centering
    \subfloat[XeonPhi]{
  \includegraphics[width=\textwidth]{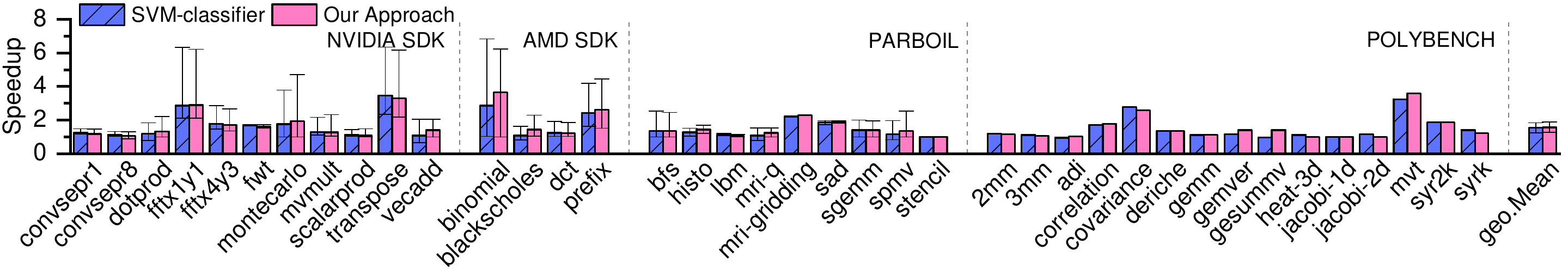}
  \label{fig_compare_svc_a}
    }\\
    \subfloat[NVIDIA GPU]{
  \includegraphics[width=\textwidth]{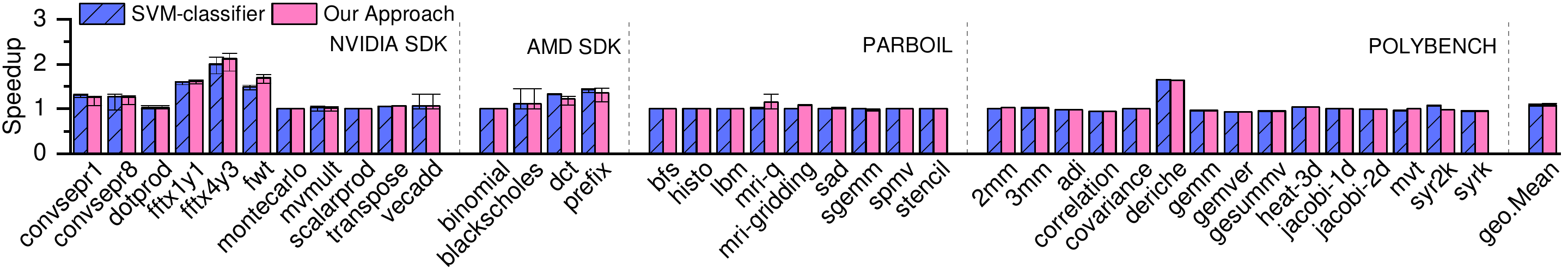}
  \label{fig_compare_svc_b}
    }
  \caption{Comparing against a classification based approach on XeonPhi (a) and NVIDIA GPU (b).
  }\label{fig_compare_svc}
  \vspace{-5mm}
\end{figure*}

Our prior work uses a \SVM classifier to predict the configurations~\cite{ipdps18}. Compared with it, the
regression-based model presented in this article has several advantages.

A classification model predicts which of a set of predefined labels the input belongs to. Using this strategy, we will need to label each
unique stream configuration. This will lead to a total of 175 labels for 311 profiling samples on the XeonPhi, and 11 labels on the GPU. On
the XeonPhi, the ratio of samples to labels is too small to build an accurate model. As a result, we have to merge labels in our prior
work~\cite{ipdps18} at the cost of losing accuracy. Classification is a constraint optimization problem where the model has to know all the
possible configurations during training. Our new regression-based approach avoids this pitfall by directly modeling the impact of the
stream configuration; it thereby can be used on any stream configuration as the configuration is the model's input.


Figure~\ref{fig_compare_svc_a} presents results obtained on the XeonPhi. Our regression-based approach outperforms the
\SVMC in 21 of the 39 programs and achieves over 5\%
 performance improvement for 13 programs.
It is to note that the overhead for ranking stream configurations is included in the experimental results. Overall, our
regression-based approach improves the \SVMC by, on average, 3\% (up to 46\%). Unlike XeonPhi, we were able to obtain
sufficient training samples per label (because the optimization space is smaller) on the GPU to build a more accurate
classification model.  As can be seen from Figure~\ref{fig_compare_svc_b}, the average speedup of \SVMC and the
regression-based approach is comparable.

Compared to a classifier, our regression-based approach has the advantage of being able to be applied to configurations
that were not seen during the training phase.  Therefore, our approach has a better generalization ability.

\subsection{Further Analysis of Performance Results}
We now take a closer look into the performance results, using XeonPhi as a case study.

\begin{figure}[t!]
  \centering
  \includegraphics[width=0.5\textwidth]{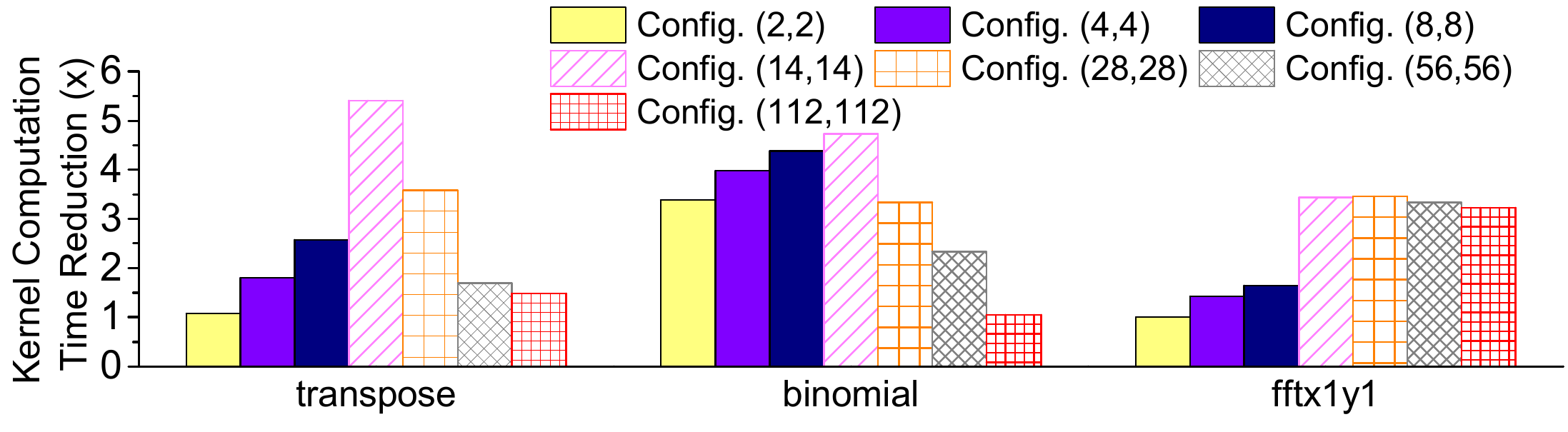}
  \caption{Reduction of kernel computation time over a single-stream execution on XeonPhi. The performance improvement comes from the reduction of the threading overhead.
  A stream configuration is annotated as (\emph{\#partitions}, \emph{\#tasks}). }
  \label{fig:extra_speedup}
\end{figure}
\subsubsection{High Speedup Cases} \label{sec:hsc} On XeonPhi, bidirectional data transfer between the host
and the accelerator cannot be overlapped, i.e., we can only issue data transfer from the host to the device or vice
versa at once but not simultaneously. As a result, the theoretical up-bound speedup for overlapping computation and
communication is 2$\times$, when the computation is perfectly overlapped with the data transfer time.
%
It is interesting to observe that several benchmarks achieve a speedup of over 2$\times$ on XeonPhi (see
Figure~\ref{fig_regression_perf}a). After having a closer investigation, we notice that such performance is attributed
to the reduction in the kernel execution time in additional to the overlapping of communication and computation.

To quantify the benefit of kernel time reduction, we measure the kernel execution time with and without multiple streams and calculate the
speedup between them. Note that we \emph{exclude the host-device communication time in this case} to isolate the contributing factors. The
kernel time improvement for \texttt{transpose}, \texttt{binomial}, and \texttt{fftx1y1} is shown in Figure~\ref{fig:extra_speedup}. As can
be seen from the diagram, choosing a good stream configuration can lead to more than 4x reduction on the kernel execution time. This is
because these benchmarks are implemented by parallelizing the inner loop within a nested loop. During runtime, the parallel threads working
on the inner loop will be created, synchronized, or destroyed for each outer loop iteration. Such threading overhead could be significant
when the outer loop iterates a large number of times. With multiple streams, we divide the whole outer loop iteration space into multiple
smaller iterations. This allows multiple groups of threads to be managed simultaneously, leading to a significant decrease in threading
overhead and faster kernel execution time. On the other hand, using too many streams and partitions will lead to a performance decrease.
This is because stream management also comes at a cost, which increases as the  number of partitions grows. Nonetheless, for applications
where the kernel computation dominates the program execution time, by reducing the kernel time can lead to additional improvement, yielding
more than 2x speedups.

\begin{figure*}[t!]
  \centering
    \subfloat[XeonPhi]{
  \includegraphics[width=\textwidth]{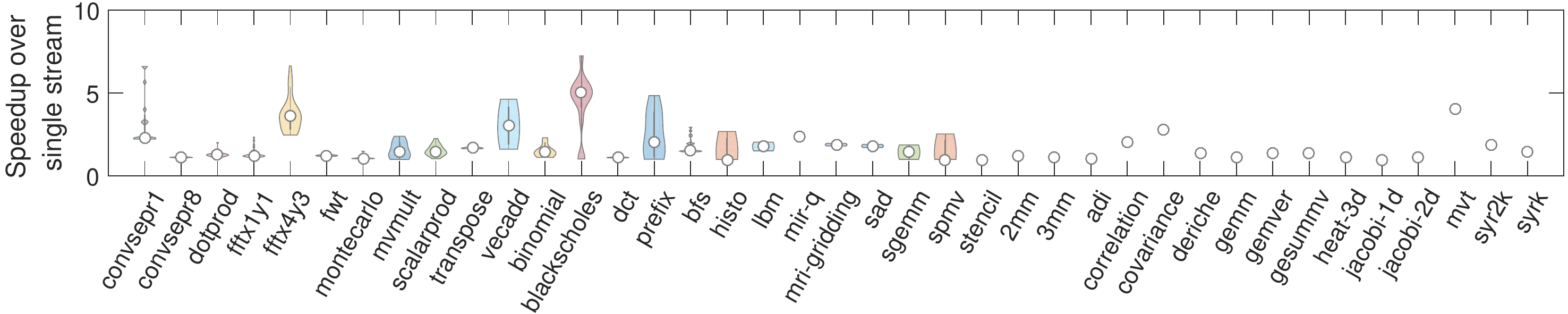}
    }\\
    \subfloat[NVIDIA GPU]{
  \includegraphics[width=\textwidth]{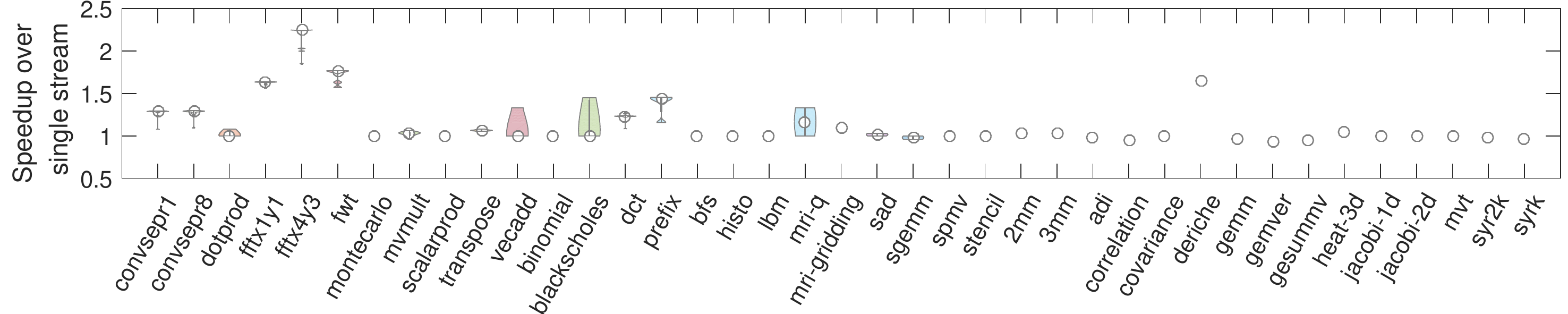}
    }
  \vspace{-2mm}
  \caption{Violin plot showing the distribution of speedups per benchmark across datasets on XeonPhi (a) and NVIDIA GPU (b).
  The shape of the violin corresponds to the speedup distribution. The thick black line shows where 50\% of the
    data lies.
  }
  \label{fig:individualdist}
\end{figure*}

\subsubsection{Speedup Distribution} Figure~\ref{fig:individualdist} gives the speedup per benchmark across datasets on
XeonPhi and the GPU. The shape of the violin plot corresponds to the speedup distribution.

On XeonPhi, we see that the speedups of \texttt{montecarlo} and \texttt{prefix} distribute fairly uniformly while the
data distribution of \texttt{fftx1y1} and \texttt{fftx4y3} is multimodal (i.e. it has two peaks). Further, the input
datasets have little impact on the behavior of \texttt{fwt} and \texttt{lbm}, so the speedups remain constant across
datasets.

On the GPU,
the speedups of \texttt{dotprod}, \texttt{vecadd}, \texttt{blackscholes} and \texttt{mri-q} distribute fairly uniformly while the data distribution
of \texttt{convsepr1}, \texttt{convsepr8}, \texttt{fftx1y1}, \texttt{fftx4y3} and \texttt{dct} is unimodal (i.e. it has one peak).
Furthermore, the input datasets have a very slight impact on the performance behaviors of \texttt{montecarlo}, \texttt{scalarprod}, \texttt{transpose} and \texttt{binomial}.
Thus, their speedups remain constant across datasets.

To conclude, the streaming speedups of some applications are sensitive to their input datasets whereas the others are not.
And the distribution of speedups on the GPU is more concentrated than XeonPhi.
This is because the current GPU implementation does not support processor core partition, the kernel execution time benefits less from multiple streams than XeonPhi.



\begin{figure}[!t]
  \centering
  \includegraphics[width=0.5\textwidth]{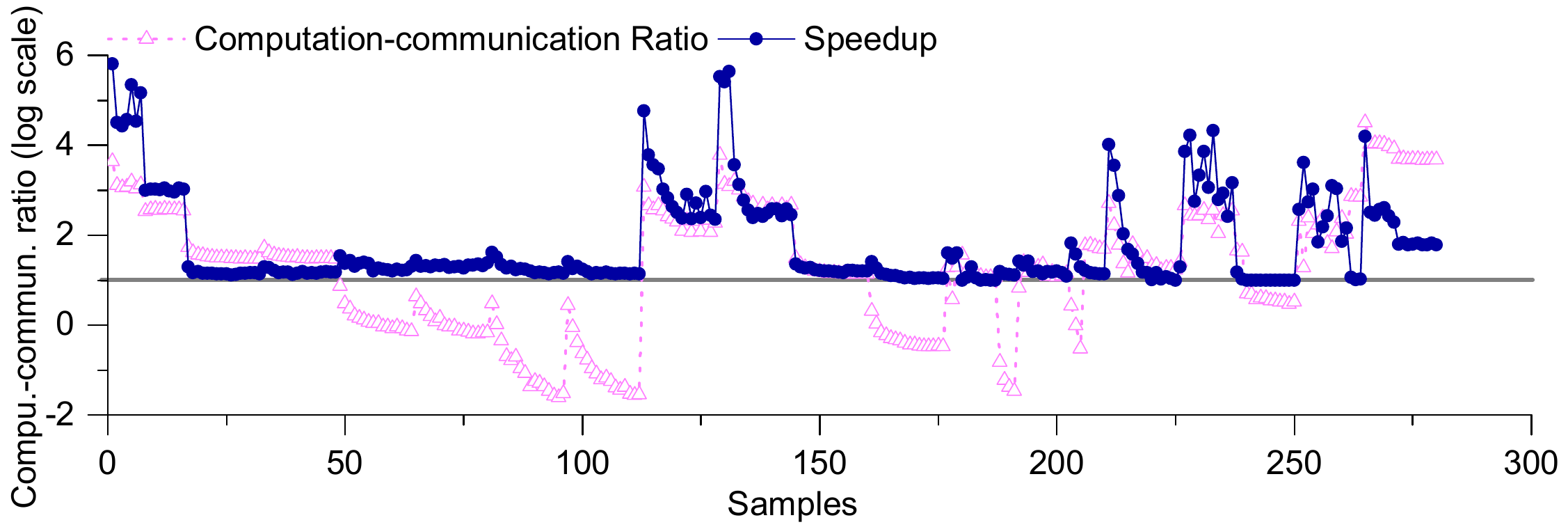}\\
  \vspace{-2mm}
  \caption{The relation between computation-communication ratio and the speedup. The computation-communication ratio is normalized using the natural logarithm function. Thus, the kernel computation time equals the host-device communication time when $ratio=0$.
  In general, a higher computation-communication ratio leads to a better speedup.
  }
  \label{fig_overall_ratio_sp}
\end{figure}

\subsubsection{Correlation Analysis} Figure~\ref{fig_overall_ratio_sp} shows the relation between the
computation-communication ratio and the achieved speedup when using heterogeneous streams across all benchmarks and
datasets on XeonPhi. We see that the computation-communication ratio varies over the benchmarks and the speedup changes
accordingly, but in general, a higher computation-to-communication ratio leads to a greater speedup. As explained in
Section~\ref{sec:hsc}, in addition to overlapping computation and communication, our approach can also reduce the
kernel computation time by choosing the right stream configuration. Therefore, benchmarks with a high
computation-communication ratio also benefit from a reduction in the kernel computation time.

 To quantify the
relation between the computation-communication ratio and the speedup, we calculate the Pearson correlation coefficient of the two
variables. The calculation gives a correlation coefficient of 0.7, indicating that the two variables (the computation-communication ratio
and the speedup) have a strong linear correlation. By carefully selecting the stream configuration, our approach tries to maximize the
overlap between communication and computation, which thus leads to favourable performance.

\subsubsection{Impact of Streaming Parallelism}
\begin{figure}[!t]
\centering
\begin{minipage}[b]{0.3\columnwidth}
	\centering %
	\includegraphics[width=\textwidth]{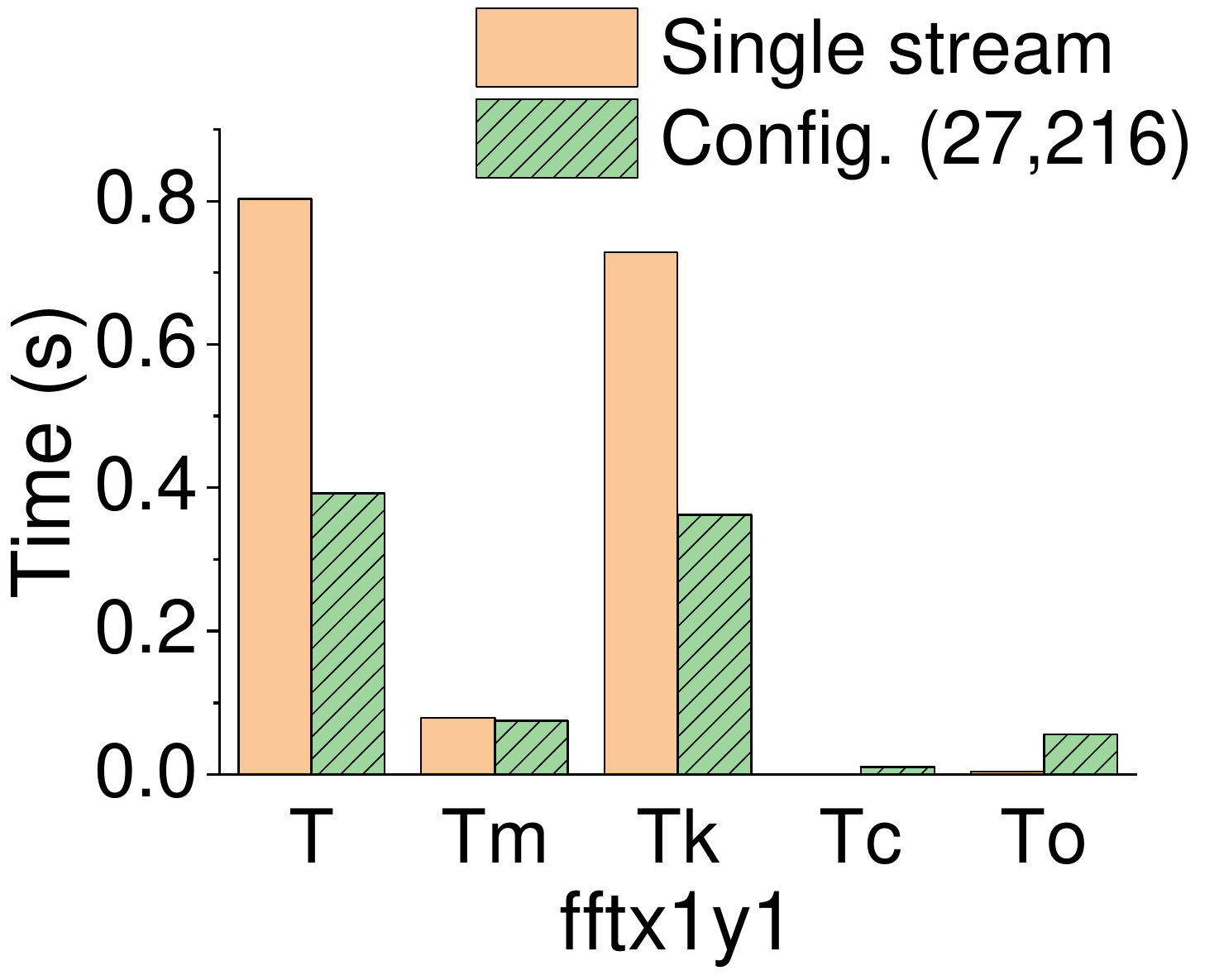}%
	\label{fig_bd_fftx1y1}
\end{minipage}
\begin{minipage}[b]{0.3\columnwidth}
	\centering %
	\includegraphics[width=\textwidth]{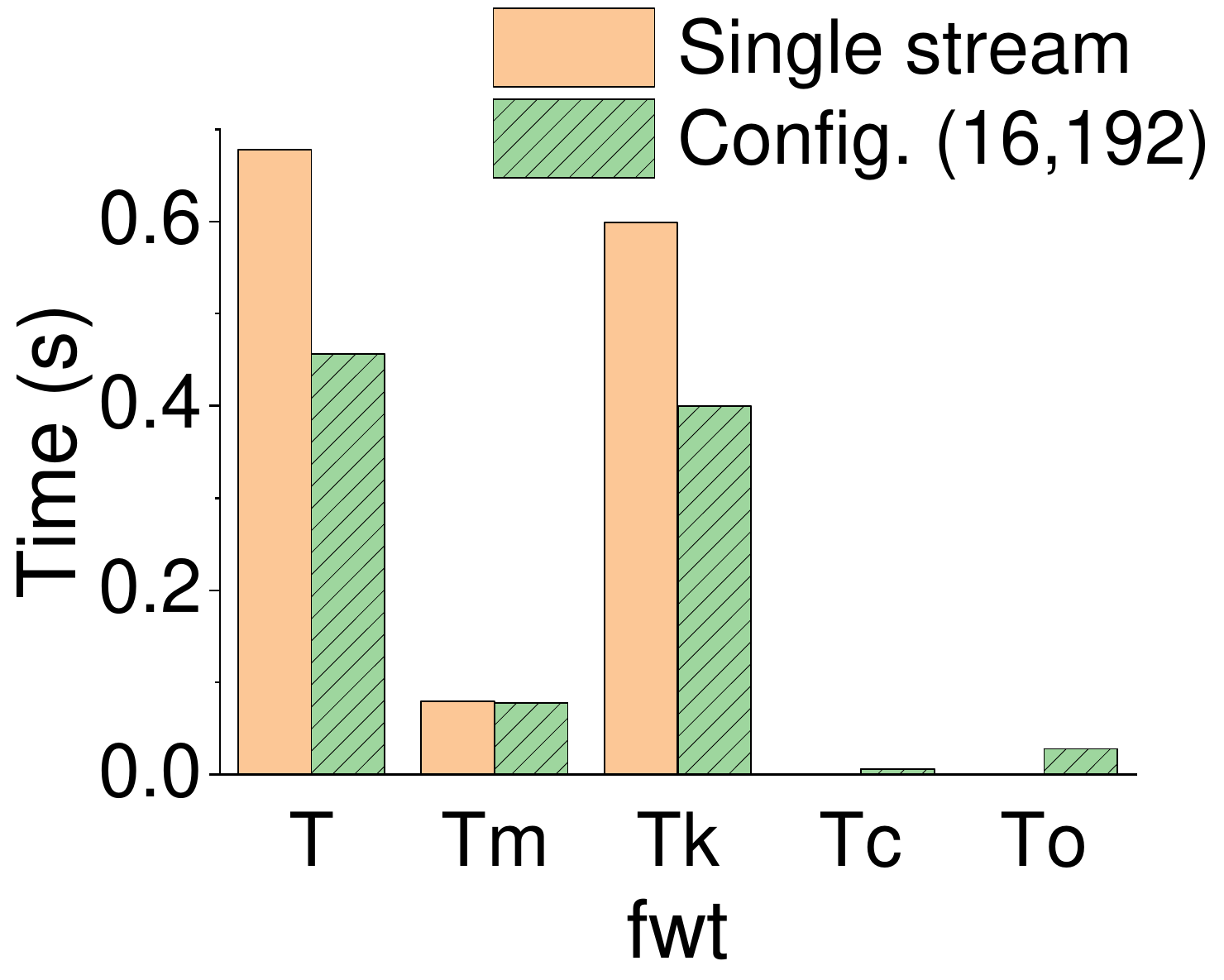}%
	\label{fig_bd_fwt}
\end{minipage}
\begin{minipage}[b]{0.3\columnwidth}
	\centering %
	\includegraphics[width=\textwidth]{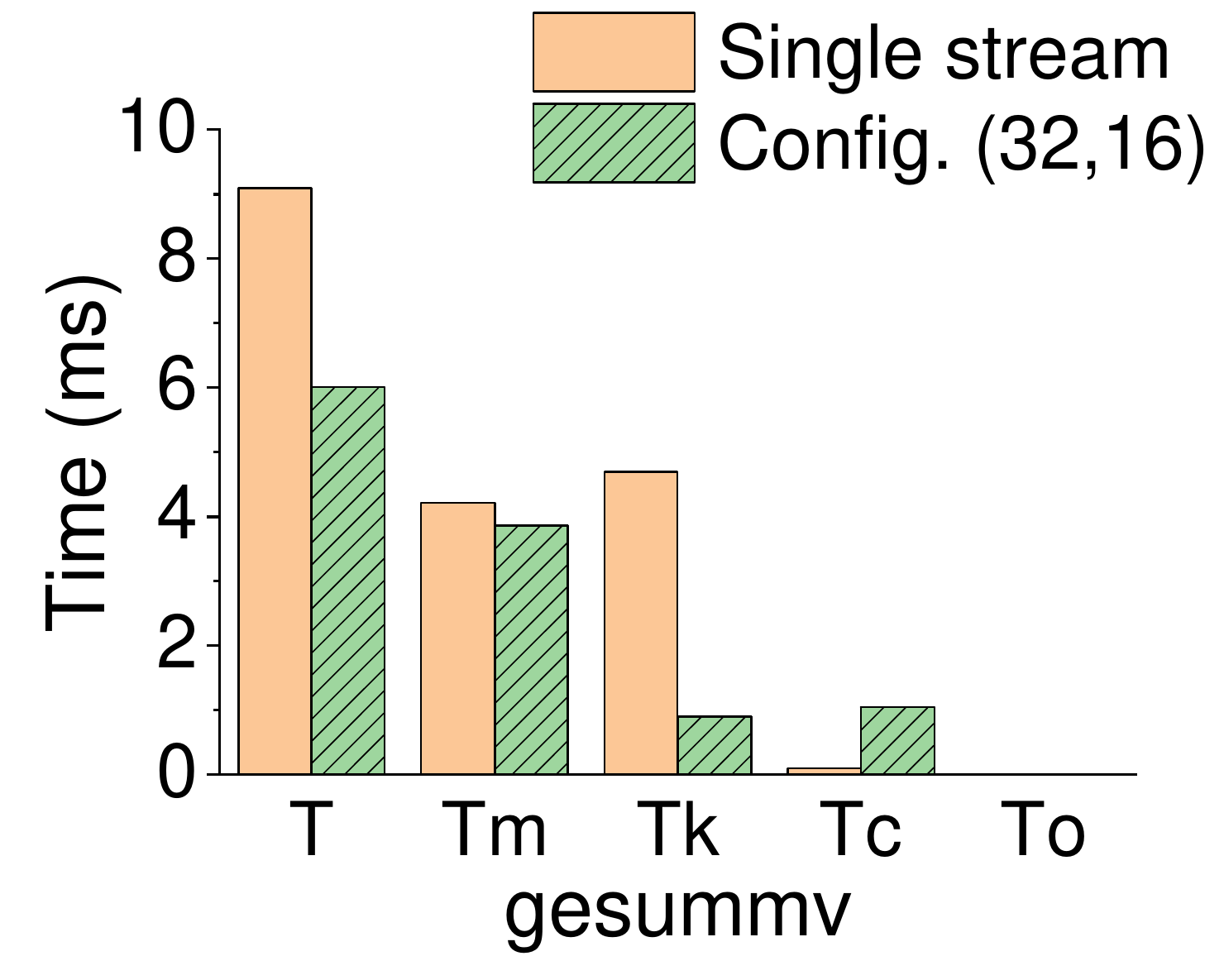}%
	\label{fig_bd_gesummv}
\end{minipage}
\caption{Breakdown of program execution time ($T$), host-device data transfer time ($T_m$), kernel execution time ($T_k$), \hStreams context initialization overhead ($T_c$) and communication-computation overlapping time ($T_o$) for single and best-performing multi-stream configurations.}
\label{fig_breakdown}
\end{figure}

Our earlier experiments show that by carefully exploiting streaming parallelism, we can significantly improve
application performance. We now take a closer look at three representative benchmarks, \texttt{fftx1y1}, \texttt{fwt}
and \texttt{gesummv}, to get a better understanding of streaming performance on XeonPhi.
These benchmarks represent different degrees of benefits obtained from streamed parallelism (with a speedup of 2$\times$,
1.5$\times$ and 1$\times$, respectively).

We use the following analytical model to breakdown the execution time of a multi-stream program:

\begin{equation}
T = T_m + T_k + T_c - T_o
\label{eq:breakdown}
\end{equation}

where $T_m$ is host-device data transfer time, $T_k$ is kernel execution time, $T_c$ is the overhead for initializing the context, and
$T_o$ is overlapping time between data transfer and kernel execution. We measure $T$, $T_m$, $T_k$, and $T_c$, and use the measurements to
calculate $T_o$.

Figure~\ref{fig_breakdown} gives the breakdown for the five components in Equation~\ref{eq:breakdown}.  For each testing program, we
compare the single-stream configuration against the best-performing multi-stream configuration. The host-device data transfer time, $T_m$,
is nearly constant among a single and a multiple stream configuration, but multi-streaming can reduce the kernel execution time, $T_k$, by
exploiting the spatial sharing of processing resources among computation tasks.  The overhead of initializing the \hStreams context, $T_c$,
depends on the kernel execution time. For \texttt{fftx1y1} and \texttt{fwt}, whose kernels run for a sufficiently long time, this one-off
runtime overhead is negligible. However, for \texttt{gesummv}, this overhead cannot be ignored due to the relatively short kernel running
time. The contribution for overlapping host-device communications with kernel execution, $T_o$, varies across programs. For
\texttt{fftx1y1} and \texttt{fwt}, it accounts for around 50\% of $T_m$, suggesting that by exploiting temporal sharing to overlap communication
with kernel execution can amortize the host-device communication overhead. For \texttt{gesummv}, $T_o$ is small due to little alignment
between data transfer and kernel execution. As such, there is little benefit for exploiting temporal sharing for this program.

This experiment gives a more detailed analysis for the benefits of exploiting multiple streams. The results reinforce our claim that the
benefit for streaming parallelism depends on the computation kernel and hence an adaptive scheme for choosing the optimal stream
configuration is needed. Our work aims to offer such a capability.

\subsection{Analysis of Predictive Modeling Techniques}

In this section, we analyse the working mechanism of our predictive model, using XeonPhi as an evaluation platform.

\subsubsection{Comparison to Alternative Modeling Techniques} \label{sec_compare_learning_techniques}
We compare our \MLP-based model against four widely used regression methods: the \DCT (Decision Tree), the \texttt{RF} (Random Forest), the \texttt{XGB} (eXtreme Gradient Boosting) and \SVM (Support
Vector Machine) as well as four classification models: \SVM, \texttt{DCT}, \MLP and \texttt{KNN} (K-Nearest Neighbors).
We use the Radial basis function kernel for the \SVM models.

For each technique, we follow the same training methodology and use the same features and training examples to build a model. For
classification models, we apply the label merging process described in our prior work~\cite{ipdps18} to improve the prediction accuracy.
Table~\ref{tbl:regression_models} compares the training overhead, average prediction time and achieved average speedup for each model. We
note that training a regression-based \SVM model has the largest overhead. Although training a \DCT has less overhead over our \MLP-based
regression model, \MLP gives better prediction performance.
The \texttt{RF} and \texttt{XGB} models are based on \DCT,
but they do not yield a better performance.
 Compared to regression models, a classification model takes less time to train
and make predictions. However, classification models give worse performance over regression models as they require more training data to
cover the optimization space. Overall, we choose to use a regression-based approach and employ \MLP because it gives the best overall
prediction performance and has modest training overhead.


\begin{table}[!t]
\scriptsize
\caption{Comparison to alternative modeling techniques}
\label{tbl:regression_models}
\centering
\begin{tabular}{lrrr}
\toprule
\textbf{Technique}  &  \textbf{Training time} &  \textbf{Avg. pred. time} & \textbf{Avg. speedup}\\
\midrule
\rowcolor{Gray} \SVM (regression) 	   &  100 hours 	& 2280 ms & 1.56 \\
\DCT (regression)	   &  65.57 seconds	& 0.74 ms & 1.51 \\
\rowcolor{Gray} \texttt{RF} (regression) 	   &  317.89 seconds 	& 11.94 ms & 1.51 \\
\texttt{XGB} (regression)	   &  28.46 seconds	& 0.74 ms & 1.49 \\
\rowcolor{Gray} \MLP (regression, ours)   &  245.8 seconds	& 0.76 ms & 1.57 \\
 \SVM (classifier)	   &  1.28 seconds	& 0.10 ms & 1.53 \\
\rowcolor{Gray}  \texttt{DCT} (classifier)    &  0.79 seconds	& 0.05 ms & 1.38 \\
\texttt{MLP}(classifier)	   &  46.45 seconds	& 0.05 ms & 1.41 \\
\rowcolor{Gray}  \texttt{KNN} (classifier)    &  0.22 seconds	& 0.23 ms & 1.43 \\
\bottomrule
\end{tabular}
\end{table}

\subsubsection{Feature Engineering\label{sec:fengineering}}
Feature engineering has a significant impact on the performance of a machine learning model (Section~\ref{sec_mlstream_modeling_features}).
Here we quantify the impact of feature engineering methods. In this work, we consider three standard feature engineering approaches
including standardization, normalization and dimension reduction.

\cparagraph{Standardization} converts all features value to be in a common range, e.g., between 0 and 1. The idea is to prevent the feature
value range to dominate the importance of that feature. In this work we apply a commonly used standardization method called
\emph{Z-score}~\cite{citeulike:0470458365} to standardize the raw feature values and the speedups (i.e., prediction targets) in the training data. We
found that feature standardization improves the achieved speedup by 3\% on average, and speedup standardization improves the  achieved
speedup by 5\% on average.

\cparagraph{Normalization} scales the feature values to make them follow the normal distribution. We have tested a range of normalization
methods including the square root, the reciprocal of square root and the natural logarithm transformation. However, we found that
normalization does not improve our model prediction accuracy.

\cparagraph{Dimension reduction} reduces the number of features, which is often useful when the number of training examples is not
proportional to the number of feature dimensions. In this work, we apply factor analysis (\FA)~\cite{citeulike:1138831995} and principal component
analysis (\PCA)~\cite{pca} to the raw features. Applying \PCA and using 9 \PCA components gives the best overall result, by improving
the average speedup by 17\%. \PCA outperforms \FA which gives an average 3\% improvement on the achieved speedup.

\subsubsection{MLP Parameters Tuning\label{sec:ptuning}}
We now discuss the impact of the \MLP parameter choices. There are four configurable parameters for an \MLP model: the activation function,
the number of hidden layers, the number of neurons, and the learning algorithm (i.e., the solver). For activation functions, we consider
\texttt{identity}, \texttt{logistic}, \texttt{tanh} and \texttt{relu}. For hidden layers and neurons, we vary the number of hidden layers
from 1 to 5 and the number of neurons per layer from 3 to 100. For the solver, we consider three commonly used weight optimizers:
\texttt{lbfgs}, \texttt{sgd} and and \texttt{adam}. We use scikit-learn implementations for the activation function and the solver. Our
experimental results suggest that the best-performing activation function and solver are \texttt{tanh} and \texttt{adam} respectively, and
using three hidden layers with 9 neurons per layers gives the best overall results on our training data. Overall, tuning \MLP model
parameter improves the average speedup by 5\% over the default parameter setting.

\begin{figure*}[t!]
  \centering
  \includegraphics[width=0.85\textwidth]{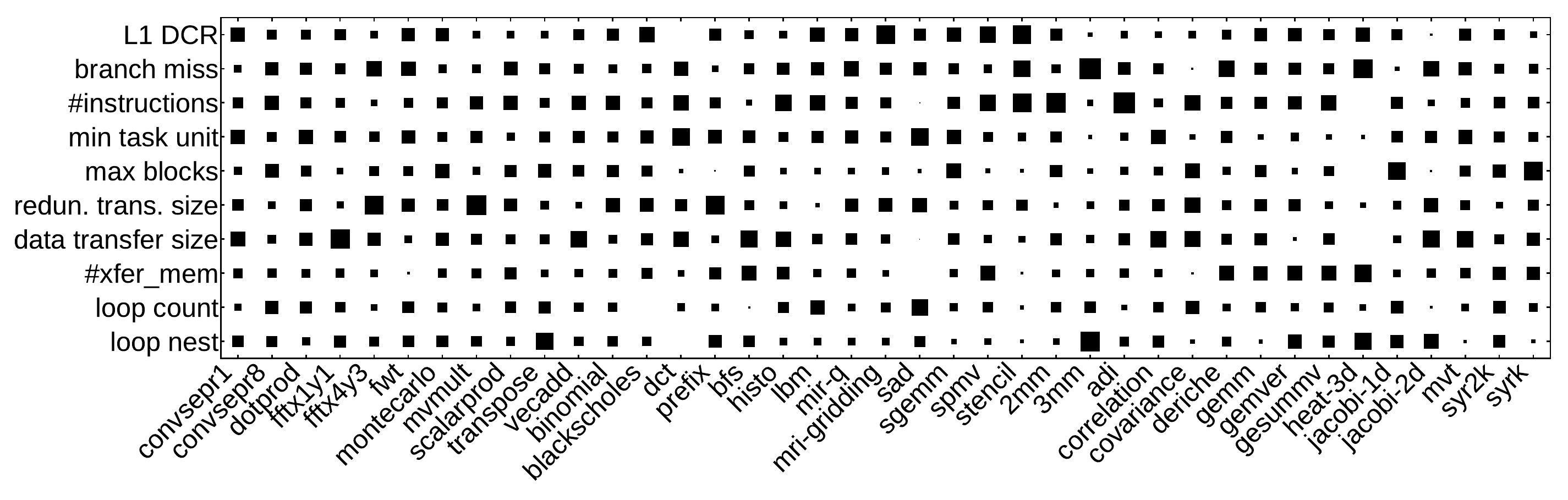}\\
  \vspace{-2mm}
  \caption{A Hinton diagram showing the impact of each feature used by the performance model to the resultant application performance. The larger the box, the more likely a feature has a greater impact on the performance of the respective benchmark.}
  \vspace{-3mm}
  \label{fig:hinton}
\end{figure*}

\subsubsection{Impact of Individual Feature\label{sec:fimp}}
In this experiment, we consider the impact of a specific feature to the resultant performance. Figure~\ref{fig:hinton} presents a Hinton
diagram illustrating how important a feature contribution to the performance model prediction accuracy (which in turns affects the
resulting application performance). The larger the box, the more significant a feature for a given program's performance. Here, the x-axis
denotes the programs, and the y-axis denotes the features used by our performance model. The impact of a feature is quantified by measuring
how much speedup improvement can be obtained if that feature is used by the performance model. Note that this is a post-hoc analysis and,
in general, we cannot know in advance the importance of a feature on \emph{unseen} programs. Figure~\ref{fig:hinton} shows that all the
features are important for the set of benchmarks targeted in the work, but the importance of features varies across programs. This diagram
illustrates how hard it is to develop an analytical model to capture the diverse behaviors and characteristics of streaming programs.

\vspace{-4mm}
\section{Related Work} \label{sec_mlstream_relatedwork}

Our work builds upon the following past foundation, while qualitatively differing from each.

\cparagraph{Task Scheduling.} There is considerable work on distributing work across heterogeneous processors to improve application
performance~\cite{citeulike:13920415, Luk:2009:QEP:1669112.1669121, shen2016workload}.
Prior work in the area typically assumes that the processor configuration is fixed and relies on the operating system to schedule parallel
tasks across processing units. Recent studies show that by partitioning the processing units into groups it is possible to significantly
improve the application performance by overlapping the host-device communication and computation on coprocessors like Intel XeonPhi~\cite{DBLP:journals/ppl/FangZLTCCY16,DBLP:conf/ipps/NewburnBWCPDSBL16}. However, existing approaches rely on static tuning to find the
processor partition and the best number of streams to run within a partition. As a result, previous approaches cannot adapt to the change
of program inputs. As a departure from prior work, we develop an automatic approach to dynamically adjust the processor partition and
task-granularity during runtime, considering the characteristics of applications and input datasets; our approach thereby can adapt to the
change of program inputs.

\cparagraph{Domain-specific Optimizations} There is considerable work on domain-specific optimization on Intel XeonPhi.
Cheng \etal~\cite{cheng2018many} and Jha \etal~\cite{jha2015improving} show that in-memory database applications suffer
from under-utilization of processor resources and hence a fine-grained tuning approach is required. Mrphi is a
framework for optimizing MapReduce workload on the XeonPhi~\cite{lu2015mrphi}. It employs a set of techniques to improve
the resource utilization to obtain higher application performance. Other works look at performance optimization for
numerical solvers~\cite{lastovetsky2017model}, sparse matrix vector multiplication~\cite{tang2015optimizing,
guney2017optimizing}, and dynamic stochastic economic models~\cite{tang2015optimizing}.
Ferr{\~a}o~\etal~\cite{ferrao2018stream} and Memeti~\etal~\cite{memeti2018hstream} develop a stream processing framework
for  XeonPhi to increase the programming productivity. The runtime can automatically distribute workloads across
CPUs and accelerating devices. These approaches improve the processor utilization by adjusting the algorithmic design,
which are complementary to our work on tuning multi-streaming parallelism for data parallel applications.

\cparagraph{Multiple Streams Modeling.} Gomez-Luna \etal~\cite{citeulike:9715521} develop a set of models to estimate the asynchronous data
transfer overhead on different GPU architectures. The models can be used to estimate the optimal number of streams to use on a given GPU
platform. Werkhoven \etal~\cite{citeulike:13920334} present an analytical model to determine  when to apply an overlapping method on GPUs.
Liu \etal~\cite{citeulike:13920353} also develop an analytical based approach to determine the optimal number of streams to use on GPUs.
However, none of these approaches considers the processor partition. As we have shown in Section~\ref{sec:alt}, ignoring the processor
partitioning parameter can lead to poor performance on Intel XeonPhi. Furthermore, these hand-crafted models have the drawback of being
not portable across architectures as the model is tightly coupled to a specific GPU architecture. Our work advances prior work by employing
machine learning to automatically learn the optimal processor partition and the number of streams/tasks to use. Since our models are
automatically learned from empirical observations, one can easily re-learn a model for a new architecture.

\cparagraph{Predictive Modeling.} Recent studies have shown that machine learning based predictive modeling is effective in code
optimization~\cite{cummins2017end,mlcpieee}, performance predicting~\cite{zhao2016predicting,wang2013using}, parallelism
mapping~\cite{Tournavitis:2009:THA:1542476.1542496,wang2010partitioning,DBLP:journals/taco/WangGO14,wang2014integrating,taylor2017adaptive},
and task scheduling~\cite{emani2013smart,marco2017improving,ren2017optimise,conext18,yuan2019using,taylor2018adaptive}. Its great advantage
is its ability to adapt to the ever-changing platforms as it has no prior assumption about their behavior. The work presented by Wen
\etal~\cite{wen2014smart} employs \SVMs to develop a binary classifier to predict that if a given OpenCL kernel can achieve a large speedup
or not. Our work differs from~\cite{wen2014smart} in that it targets a different architecture and programming model, and it predicts from a
larger number of configurations instead of making a binary prediction. Our prior work developed an \SVM based classifier to predict the
optimal stream configuration for Intel XeonPhi~\cite{ipdps18}. However, it requires having sufficient training data samples to cover all
possible stream configurations. Our approach improves the prior work by directly modeling the impact of the stream configuration. As a
result, our approach can make predictions for any stream configuration (even those are not seen in the training data).

\cparagraph{Autotuning Parallel Programs.} Our approach is closely related to autotuning that searches for the best-performing optimization
configuration~\cite{datta2008stencil,ansel2014opentuner}. This technique is demonstrated to be effective for choosing algorithmic
choices~\cite{halide}, tuning GPU code~\cite{nukada2009auto,tillet2017input, dao2018auto}, optimizing structured parallel
programs~\cite{dastgeer2011auto,thiagarajan2018bootstrapping,chen2019optimizing} and non-uniform memory access (NUMA)
architectures~\cite{katagiri2017auto}, and more recently for deep neural networks~\cite{liao2018uhcl}. Many of the prior works in this area
employ an evolutionary-based approach by applying and profiling candidate optimization options to choose a good option to use. One of the
key changes of autotuning is how to avoid the profiling overhead which could be prohibitively expensive. We do so by using a performance
model to quickly evaluate the profitability of a candidate optimization option. We show that our approach has low runtime overhead, which
thus permits us to apply it at runtime to best match the optimization strategy to the program input.  Furthermore, our work is the first
for tuning heterogeneous streaming parallelism on heterogeneous many-cores (XeonPhis and GPUs).

\cparagraph{Automatic Generation of Parallel Programs.} The OpenMPC compiler~\cite{DBLP:conf/sc/Lee2010} translates OpenMP to CUDA
programs. Wang \etal~\cite{grewe2013portable,DBLP:journals/taco/WangGO14,cc14} translates OpenMP to OpenCL programs and use machine
learning to select the most suitable device from the host CPU and the GPU to run the code. Rawat \etal presents an automatic approach to
generate GPU code from a domain-specific language (DSL) for stencil programs~\cite{8451874}. All of the above approaches target GPUs, and
do not utilize the multi-streaming strategy.

\vspace{-2mm}
\section{Conclusion} \label{sec_mlstream_conclusion}
This article has presented an automatic approach to exploit streaming parallelism
on heterogeneous many-cores. Central to our approach is a machine learning-based model that predicts the resulting performance when running
the target application under a given streamed configuration. The performance predictor is then used as a cost function to quickly rank
candidate configurations at runtime, to determine which stream configuration should be used on a per-program per-dataset basis. We have
evaluated our approach on an Intel XeonPhi and an NVIDIA GTX 1080 Ti GPU, with 39 representative benchmarks.  Experimental results show
that our approach delivers an average speedup of 1.6x and 1.1x on XeonPhi and the GPU, respectively. These results translate to over 93\%
of the best-available performance.


  \section*{Acknowledgment}
This work was partially funded by the National Key Research and Development Program of China
under Grant No. 2018YFB0204301, the National Natural Science
Foundation of China under Grant agreements 61972408, 61602501 and 61872294;
 For any correspondence, please contact Jianbin Fang (Email: j.fang@nudt.edu.cn).
\ifCLASSOPTIONcaptionsoff
  \newpage
\fi


\bibliographystyle{IEEEtran}
\bibliography{IEEEabrv,mybibs}
\vspace{-12mm}


\end{document}